\documentclass[12pt,preprint]{aastex}
\shorttitle{Maximum Entropy and SN Parameters}
\shortauthors{Summerscales et al.}

\begin{document}

\title{Maximum Entropy for Gravitational Wave Data Analysis: Inferring the
Physical Parameters of Core-Collapse Supernovae}

\author{T. Z. Summerscales\altaffilmark{1}}
\affil{Center for Gravitational Wave Physics, Penn State University,
  University Park, PA 16802}
\email{tzs@gravity.psu.edu}
\altaffiltext{1}{Now at Department of Physics, Andrews University,
  Berrien Springs, MI, 49104} 
  
\author{Adam Burrows}
\affil{Steward Observatory, University of Arizona, 933 North Cherry
  Avenue, Tucson, AZ 85721}
\email{burrows@zenith.as.arizona.edu}

\author{Lee Samuel Finn}
\affil{Center for Gravitational Wave Physics, Penn State University,
  University Park, PA 16802}
\email{lsfinn@psu.edu}

\and

\author{Christian D. Ott}
\affil{Steward Observatory, University of Arizona, 933 North Cherry
  Avenue, Tucson, AZ 85721}
\email{cott@as.arizona.edu}


\begin{abstract}
  The gravitational wave signal arising from the collapsing iron core
  of a Type II supernova progenitor star carries with it the imprint
  of the progenitor's mass, rotation rate, degree of differential
  rotation, and the bounce depth.  Here, we show how to infer the
  gravitational radiation waveform of a core collapse event from noisy
  observations in a network of two or more LIGO-like gravitational
  wave detectors and, from the recovered signal, constrain these
  source properties. Using these techniques, predictions from recent
  core collapse modeling efforts, and the LIGO performance during its S4
  science run, we also show that gravitational wave observations by
  LIGO might have been sufficient to provide reasonable estimates of
  the progenitor mass, angular momentum and differential angular
  momentum, and depth of the core at bounce, for a rotating core collapse
  event at a distance of a few kpc.
\end{abstract}


\section{INTRODUCTION}

The gravitational waves that we expect to observe in large detectors
such as LIGO \citep{waldman:2006:sol,mandic:2006:sol}, Virgo
\citep{acernese:2006:vs}, GEO600 \citep{luck:2006:sog},
TAMA300~\citep{takahashi:2004:sot} and LISA
\citep{robertson:2000:trl,rudiger:2004:l-l}, reflect the coherent
evolution of the most compact part of the source mass distribution.
From the observed waves we have the potential to infer the factors that
govern that evolution. For example, the evolution of the collapsing stellar
core in a type II supernova is determined in part by the progenitor mass
density and angular momentum distributions in the inner core.
None of these properties of the progenitor can be directly
determined from the electromagnetic radiation we observe when the
shock emerges from the stellar envelope. In addition, the forces that
govern the evolution of the core depend upon many things that are
unknown, or poorly modeled, today (for example, the matter
equation of state, the role played by neutrinos and neutrino 
radiation transport, general relativity, convection and
non-axisymmetry). The gravitational waves emitted
during the collapse phase and its aftermath carry the signature of all these
preconditions and the dynamics that govern both the collapse and
subsequent rebound.  Reading that signature requires inferring the
waveform from the noisy detector observations. 
Here, we develop a new method for inferring the gravitational radiation 
waveform from the  noisy data from two or more detectors, based on 
Jaynes' principle of maximum entropy \citep{jaynes:1957:ita,jaynes:1957:ita:1}, 
demonstrate its effectiveness when applied to the discovery of signals arising 
from simulated rotating iron core collapse buried in simulated detector 
noise, and show how the inferred waveform can be used, in principle, to 
gain insight into the properties of the source.

The first problem addressed in this study is the problem of inferring
the gravitational wave signal from the data produced by the detectors,
also referred to as the deconvolution or inverse problem. Inverse
problems generally have been long-recognized as a problem to be
approached with great care (see \citealt{evans:2002:ipa} for a recent
review). The detector response, which relates the incident
gravitational wave to the signal observed in the detector, is
generally an ill-conditioned function; additionally, the presence of
additive noise generally confuses the observation.  Together, the
noise and the ill-conditioned detector response generally thwart
na\"{\i}ve attempts at signal deconvolution.  Early attempts at
gravitational wave signal deconvolution explored a least-squares or
maximum likelihood approach to deconvolution: e.g.,
\citet{gursel:1989:nos} developed a procedure for inferring a plane
wave signal incident on a network of detectors with a
frequency-independent response, in which case the problem response is
well-conditioned.  These techniques tend to over-fit the observations;
additionally, realistic detectors have frequency responses that are
ill-conditioned and this complicates a least-squares or maximum
likelihood approach to deconvolution. More recent efforts have
explored regularized methods for inferring the incident gravitational
radiation waveform \citep{rakhmanov:2006:rda}.  Here, we use the
\emph{maximum entropy} principle to regularize the deconvolution problem,
developing an application to the gravitational wave inference problem
that is applicable in the general case of a frequency dependent and
ill-conditioned detector response and avoids over-fitting the
observations in the presence of additive detector noise. Going beyond
the problem of inferring the incident wave, we use the inferred
waveform to explore how, and how well gravitational wave
observations of this kind can be used to learn about the source.

Maximum entropy approaches to deconvolution have a long heritage in
astronomical image reconstruction~\citep{%
ponsonby:1973:emf,%
gull:1978:irf,%
gull:1979:mem,%
skilling:1984:mei,%
steenstrup:1985:dip,%
shevgaonkar:1987:mem,
nityananda:1982:mei,%
nityananda:1983:rop,%
narayan:1984:me-,%
narayan:1986:mei,%
pantin:1996:doa,%
starck:1996:mme,%
starck:1997:air,
barreiro:2001:rms,%
maisinger:2004:mir%
}, where they have been used in all wavebands. Recent examples of the
use of maximum entropy based methods include the reconstruction of the
cosmic microwave
background~\citep{%
maisinger:1997:mem,%
vielva:2001:cma%
} including maps based on data from WMAP~\citep{bennett:2003:fwm} and
COBE~\citep{%
jones:1998:1tc,%
jones:1999:mmj,%
barreiro:2004:fsu%
}.\footnote{In all these applications the Principle of Maximum Entropy actually plays
a relatively small role: much more important is Bayes Law and the understanding 
that probability can represent degree of belief or certainty in the state of a system;
nevertheless, the term Maximum Entropy has come to refer to all these methods
of deconvolution and we continue that tradition.} 
Our work applies that heritage to the simpler problem of
reconstructing the time-dependent plane wave signal incident on a
network of gravitational wave detectors from their time series
response.

Once we have inferred the incident signal, we are faced with a second
problem of inferrence: identifying the properties of the source from
the signal.  One way of associating gravitational waveforms with
supernova properties is to compare the inferred waveform with models
arising from simulations that explore the signal dependency over a
broad range of physical parameters. We should expect the inferred
waveform to have the most in common with the simulated waveforms
arising from models whose character is most similar to the actual
source.  Here, we compare, using the cross-correlation, inferred
waveforms with simulated waveforms arising from different rotating
core collapse models. We assume that models whose signal shows the greatest
correlation are the ones most likely to be similar to the
source.\footnote{Note the distinction between correlation and the more
  involved matched filtering \citep{finn:1992:dma,finn:1993:obi}.
  Matched filtering is a useful analysis technique when the functional
  form of the signal being sought is known precisely. Here, we presume
  that we have only a qualitative model of the signal dependence on
  the physical parameters of interest. In that case, using the full
  apparatus of matched filtering could very well lead us to reject
  real signals because the match to the model is only qualitative.}

Section~\ref{sec:maxent} describes in detail the development of the inference Bayesian inference method that we have developed to identify the time-dependent gravitational wave
signal incident on a network of detectors, ending with a demonstration of the method applied to simulated observations made at an idealized two-detector network. 
Section~\ref{sec:Application} reviews the expected gravitational-wave
emission processes in core-collapse supernovae and  describes the  
recently produced catalog of rotating core-collapse waveforms
by \citet{ott:2004:gwf} and the physics that went into the considered
models. We use these simulated core-collapse signals to characterize 
how well our procedure for inferring the incident signal and 
characterizing the source works.
Finally, in \S~\ref{sec:Conclude} we summarize our
conclusions and directions for further study.

One important goal with this paper is to connect the two communities of 
gravitational wave experimentalists and supernova modelers in a way 
that has not been done in the past.  For this reason we have taken particular
care to make the discussion of \S~\ref{sec:maxent} pedagogical in nature. 
Future gravitational wave models 
can be put through the pipeline established with this paper so that we 
can obtain more credible estimates of what might be possible with either 
initial or advanced LIGO, Virgo, GEO600, or TAMA300, or with any 
combination of these, for any theoretical model.  This has simply never 
been done for any collapse models, and is a novelty of this paper that 
we hope will stimulate further interactions between astrophysicists doing 
supernova simulations and the gravitational wave detection and data-analysis 
communities.

\section{A Bayesian Approach to Deconvolution}
\label{sec:maxent}

\subsection{Introduction}
The first problem faced in the analysis and interpretation of astronomical data is the identification of a signal in noisy observations. When the signal being sought takes a known form then special techniques, tuned to the characteristics of the signal, may be used; however, the general problem remains the same: the first requirement is to identify the characteristics of the radiation $\mathbf{h}$ incident on the detector(s) that gives rise to the observational data $\mathbf{d}$. The data may be from a radio antenna, a spectrograph, or a network of gravitational wave detectors; our goal may be to map of the radio emission from a distant AGN, the emission spectra of an accretion disk, or the burst of gravitational waves associated with a core-collapse supernova. In the case of gravitational wave data, which is our principle interest in this paper, the data $\mathbf{d}$ are a set of time series corresponding to the sampled output of each detector, one time series from each detector in our network; the radiation $\mathbf{h}$ is the time dependent gravitational wave strain incident as a plane wave on the detector(s), which may be characterized (for example) by the direction of wave propagation and the time dependent strain amplitudes $h_{+}(t)$ and $h_{\times}(t)$ in the two polarization states as measured at the Earth's barycenter. Restricting attention to linear detectors (which is almost always the case and thus hardly a restriction at all) the observations in each case are a linear superposition of a confounding noise and the incident radiation as filtered by the detector response function $\mathbf{R}$: 
\begin{equation}\label{eq:d=rh+n}
\mathbf{d} = \mathbf{R}\mathbf{h}+\mathbf{n}
\end{equation}
where $\mathbf{R}\mathbf{h}$ is the convolution of the detector network response with the incident radiation and $\mathbf{n}$ is the noise contribution to the observation $\mathbf{d}$. We generally know $\mathbf{R}$ and the statistical properties $\mathcal{N}$ of the noise $\mathbf{n}$; with these our aim is to determine $\mathbf{h}$ from $\mathbf{d}$. 

In the absence of noise the solution to this problem appears, at first blush, to be straightforward: i.e., we solve for $\mathbf{h}$ the system of equations \ref{eq:d=rh+n} with $\mathbf{n}=0$. Problems arise when $\mathbf{R}$ is not invertible, either because we have too much or two little data (i.e., the problem is over- or under-determined), or because the detector(s) is (are) insensitive to some aspects of the signal $\mathbf{h}$ (e.g., the detector cannot distinguish between polarization states of $\mathbf{h}$, or is insensitive to signal energy outside of some band). In these cases we are forced to deal with uncertainty: if the data $\mathbf{d}$ over-determine $\mathbf{h}$ we must be prepared to resolve contradictions within the data; if $\mathbf{d}$ under-determine $\mathbf{h}$ any conclusions we reach must be tempered by our incomplete knowledge of $\mathbf{h}$. 

These problems are only compounded when we consider the real case of noisy data. In real problems the response function $\mathbf{R}$ is always ill-conditioned: i.e., the response is not invertible (the problem is over-determined), or not unique (the problem is under-determined), or the solution $\mathbf{h}$ to an equation of the form $\mathbf{R}\mathbf{h}=\mathbf{d}$ is sensitive to small perturbations in $\mathbf{d}$. Sensitivity of $\mathbf{h}$ to perturbations in $\mathbf{d}$ presents a new twist: even when when the noise contribution to the observation $\mathbf{d}$ is numerically small, it is dangerous to assume that $\mathbf{R}^{-1}\mathbf{d}$ is close to $\mathbf{h}$. Finally, even in those cases where the response is invertible and well-conditioned, applying the inverse to $\mathbf{d}$ does not distinguish between the signal $\mathbf{h}$ and a ``signal equivalent noise'' $\mathbf{R}^{-1}\mathbf{n}$.  

Framing the problem of deconvolution in the language of Bayesian inference provides guidance on how to proceed in determining $\mathbf{h}$.  In a Bayesian approach to the problem of inference we seek not $\mathbf{h}$, but a probability distribution
\begin{equation}
f(\mathbf{h}'|\mathbf{d},\mathbf{R},\mathcal{N},\mathcal{I})
=
\left(
\begin{array}{l}
\mbox{probability density that $\mathbf{h}'$ is incident on the detector(s)}\\ 
\mbox{given data $\mathbf{d}$, response $\mathbf{R}$, noise
  characterization}\\ 
\mbox{$\mathcal{N}$, and other unenumerated assumptions $\mathcal{I}$.} 
\end{array}
\right).
\end{equation}
The probability density $f$, if we can find it, fully describes our legitimate knowledge --- including uncertainty --- regarding the waveform and its properties: e.g., we can compute from it our expectation of the gravitational wave power incident on the detector, etc. It also provides a good point estimate of the incident radiation: i.e., the $\mathbf{h}'$ that maximizes $f(\mathbf{h}'|\mathbf{d},\mathbf{R},\mathcal{N},\mathcal{I})$. We can report $f$ or, as is more generally the case, report some summary of the distribution $f$: e.g., the $\mathbf{h}'$ that maximizes $f$ and some suitably defined ``error bars", which summarize the degree of our uncertainty. 

Following Bayes Law, the probability distribution $f$ may be ``factored'' into the product of three other distributions:
\begin{mathletters}\label{eq:f}
\begin{equation}
f(\mathbf{h}'|\mathbf{d},\mathbf{R},\mathcal{N},\mathcal{I}) = 
\frac{g(\mathbf{d}|\mathbf{h}',\mathbf{R},\mathcal{N},\mathcal{I})q(\mathbf{h}'|\mathcal{I})}{
v(\mathbf{d}|\mathbf{R},\mathcal{N},\mathcal{I})} ,
\end{equation}
where
\begin{eqnarray}
g(\mathbf{d}|\mathbf{h}',\mathbf{R},\mathcal{N},\mathcal{I})
&=&\left(
\begin{array}{l}
\mbox{likelihood function for $\mathbf{d}$ given $\mathbf{h}'$}
\end{array}
\right)\\
q(\mathbf{h}'|\mathcal{I})
&=&\left(
\mbox{\emph{a priori} expectations regarding the incident wave}.
\right)\\
v(\mathbf{d}|\mathbf{R},\mathcal{N},\mathcal{I}) &=& \left(\begin{array}{l}
\mbox{normalization constant equal to the probability}\\
\mbox{of observing $\mathbf{d}$ given response $\mathbf{R}$ and noise $\mathcal{N}$}
\end{array}
\right)
\end{eqnarray}
\end{mathletters}%
The probability $v$ is independent of $\mathbf{h}$. We can ignore it when our only goal is to  find the $\mathbf{h}$ that maximizes $f$; however, as we will see below, $v$ plays a vitally important role in helping us choose our prior $q$. In the following subsections we describe how to compute $f$ by finding, in turn, of the probabilities $g$, $q$ and $v$. 

\subsection{The Likelihood Function $g$}
Focus attention first on the likelihood function $g(\mathbf{d}|\mathbf{h}',\mathbf{R},\mathcal{N},\mathcal{I})$. If the signal content of $\mathbf{d}$ corresponds to incident gravitational radiation $\mathbf{h}$, then $\mathbf{d}-\mathbf{R}\mathbf{h}$ is just detector noise. Correspondingly, the likelihood function $g(\mathbf{d}|\mathbf{h},\mathbf{R},\mathcal{N},\mathcal{I})$ is just the probability that $\mathbf{d}-\mathbf{R}\mathbf{h}$ is noise. How do we evaluate this probability?

Gravitational wave detectors, interferometric or acoustic, measure a continuous quantity: i.e., they are not particle detectors that count discrete events (e.g., photons).  The statistical properties of the noise in these detectors is characterized by its power- and cross-spectral density \citep{abbott:2004:dda,%
lazzarini:2002:lss,%
lazzarini:2003:lss,%
lazzarini:2004:sbs,%
lazzarini:2005:sss%
} or, equivalently, its mean and covariance \citep{kittel:1958:esp}.\footnote{In fact, the noise cross-spectral density --- equivalent to the correlation of the noise between two different detectors --- has not yet been evaluated for or played a role in the analysis of LIGO data.} The Principle of Maximum Entropy was introduced by \citet{jaynes:1957:ita,jaynes:1957:ita:1} as a means of identifying probability distributions whose ``information content'' is, in a well-defined and relevant way, consistent with this kind of a priori information but otherwise incorporates no other assumptions. It was first articulated in the context of statistical thermodynamics, where it provided a foundation for understanding the role of the Gibbs Ensemble and entropy maximization in statistical mechanics; however, it is ultimately a logical statement about statistical inference and, as such, statistical thermodynamics is just one application of the Principle of Maximum Entropy. 

The articulation of the Principle of Maximum Entropy was made possible by the development of a theory of information and a (recovered) understanding that probability is a more general concept than ``relative frequency of occurrence''. That a probability distribution embodies information is a relatively straightforward proposition to demonstrate. Consider a physical system that can be in any one of several states and suppose that we can measure certain properties $\mu, \nu, \ldots$ of the system whose values depend on --- but do not necessarily determine --- the system's state. For example, the system might be an $N$-dimensional quantum simple harmonic oscillator, whose state is determined by the $N$ quantum numbers $n_1$, $n_2$, $\ldots$, $n_N$, and the observable might be the total energy of the system, $\mu=\hbar(N/2+\sum_k n_k)$. If we have no knowledge of the system's state then it is natural that we should regard it as equally likely that the system is in any particular state: i.e., if we were to assign a number $p$,  $0\leq p\leq 1$, to represent our degree of certainty that the system's state is $\vec{n}$, with $0$ representing complete certainty that the system is not in state $\vec{n}$ and $1$ representing complete certainty it is in state $\vec{n}$, then $p$ would be the same for all $\vec{n}$. Now suppose that we measure or are told $\mu$ is equal to $\mu_0$: what does this new information tell us about the state of the system? In our example, we now know that only those states $\vec{n}$ satisfying $\sum_k n_k$ equal to $\mu_0/\hbar-N/2$ are possible but that, among these, no $\vec{n}$ is preferred over another:  i.e., the number $p$ we should ascribe to state $\vec{n}$ vanishes if $\mu(\vec{n})$ is not equal to $\mu_0$ and is equal to a constant for all other states. Given this distribution we can express our expectations regarding other properties of the system: e.g., our expectation that the property $\nu$ takes on the value $\nu_0$ is equal to the ratio of the number of states for which $(\nu,\mu)$ is equal to $(\nu_0,\mu_0)$ to the number of states for which $\mu$ is equal to $\mu_0$. These numbers $p$ both represent our degree of certainty regarding the state of the system \emph{and} satisfy exactly the same algebraic and relational laws that we require of probabilities \citep{cox:1946:pfa,cox:1961:aop}; so, we will hereafter refer to them as probabilities. 
 
In this demonstration we see that probability is more general than ``relative frequency of occurrence'', that it does not require that we invoke an ensemble or make any assumptions regarding ergodicity --- there is just one system and it is in a definite state --- and that it represents in a meaningful way our state of knowledge about the system. Modern information theory was launched in a series of papers by \citet{shannon:1948:mto,shannon:1948:mto:1} in which he gave a definite meaning to ``information content'' in the context of a probability distribution, described the properties that a measure of information should have, and showed that there exists a unique measure of the information content of a probability distribution, which is the negative of what is conventionally understood as the entropy of an ensemble. It is more common to refer to the negative of the information content as the measure of uncertainty, or (information) entropy:
\begin{equation}\label{eq:discreteEntropy}
H(p) = -\sum_k p_k\log p_k
\end{equation}
where the sum is over all states of the system.\footnote{The base of the logarithm corresponds simply to a choice of units in which to measure entropy.} For a continuous distribution $p(\vec{x})$ the summation becomes an integral and we must normalize the probability as it occurs in the logarithm by the density of states $\rho(\vec{x})$:
\begin{equation}\label{eq:continuousEntropy}
H(p) = -\int d^nx\,p(\vec{x}) \log\frac{p(\vec{x})}{\rho(\vec{x})}.
\end{equation}

Returning to our previous example, it is clear that there may be a large number of distributions $q(\vec{n})$ that are consistent with the knowledge that the observed $\mu$ is equal to $\mu_0$. Each of these distributions has an entropy $H(q)$; of these, one --- call it $\tilde{q}$ --- will have the greatest entropy. We can find that distribution by finding the $q$ such that the variation $\delta H(q)/\delta q$, subject to the constraint that $\sum_k q_k=1$, vanishes:
\begin{mathletters}
\begin{eqnarray}
0 &=& \delta \left[H(q) + \lambda_0 \sum_k q_k\right]\nonumber\\
&=& \sum_k\left[-1 - \log q_k + \lambda_0\right] \delta q_k\\
q_k &=& \exp\left(\lambda_0-1\right)
\end{eqnarray}
\end{mathletters}%
where the Lagrange multiplier $\lambda_0$ is chosen so that $\sum_k q_k=1$. The maximum entropy distribution $\tilde{q}$ assigns equal probability to every state $\vec{n}$ such that $\mu(\vec{n})$ is equal to $\mu_0$: i.e., $\tilde{q}$ is the distribution that embodies what we understand to be just the information that $\mu(\vec{n})$ is equal to $\mu_0$, but no more. The Principle of Maximum Entropy states that this is always so: i.e., given a system with possible states $x$ and constraints on the system in the form of functions of the state, the probability distribution $p(x)$ that embodies just the information in those constraints is equal to the distribution that maximizes the entropy (either equation \ref{eq:discreteEntropy} or \ref{eq:continuousEntropy}) subject to the constraints. 

We can use the Maximum Entropy Principle to find the likelihood function $g$ --- i.e., the probability distribution that $\mathbf{d}-\mathbf{R}\mathbf{h}$ is equal to detector noise--- given the characterization of the gravitational wave detector noise described by its mean $\mu$ and autocorrelation function $c_\ell$:  
\begin{mathletters} \label{eq:meanAutocorr}
\begin{eqnarray}
\mu &=& \left<x\right> \sim \lim_{N\rightarrow\infty}N^{-1}\sum_{k=0}^{N-1}x_k\\
c_\ell &=& \left<x_k x_{k+\ell}\right> \sim \lim_{N\rightarrow\infty}N^{-1}\sum_{k=0}
\left(x_k-\mu\right)\left(x_{k+\ell}-\mu\right).
\end{eqnarray}
\end{mathletters}%
We can construct many example probability densities $p(\mathbf{x})$ whose moments take on these values. From among these we desire the particular probability distribution $\tilde{p}(\vec{x})$ that has maximum entropy: i.e., we desire the distribution that has minimum information subject to the constraints that its moments satisfy equations \ref{eq:meanAutocorr}. This distribution satisfies the variational equation
\begin{eqnarray}\label{eq:noiseDist}
0 &=& \delta\left[
-\sum_{\vec{x}} \tilde{p}(\vec{x})\log\tilde{p}(\vec{x}) - 
\lambda_0\left(1-\sum_{\vec{x}}\tilde{p}(\vec{x})\right)- 
\lambda_1\left(\mu-\sum_{\vec{x}}x_k\tilde{p}(x)\right)  \right.\nonumber\\
&&\left.\qquad{}- 
\sum_{\ell}\lambda_{2,\ell}\left(c_{\ell}-\sum_{\vec{x}}(x_k-\mu)(x_{k+\ell}-\mu)\tilde{p}(\vec{x})\right)
\right]
\end{eqnarray}
where $\lambda_0$, $\lambda_1$ and $\lambda_{2,\ell}$ are Lagrange coefficients, which are chosen to insure that the probability is normalized ($\lambda_0$), that the mean $<x_k>$ is $\mu$ ($\lambda_1$), and that the auto-correlation $<x_kx_{k+\ell}>$ takes on the value $c_{\ell}$ ($\lambda_{2,\ell}$). 
Solving equation \ref{eq:noiseDist} is straightforward: taking the variation we find
\begin{equation}
0 = \sum_{\vec{x}}\left[
\left(1+\log\tilde{f}(\vec{x})\right) + \lambda_0 + \lambda_1 x_k + \sum_{\ell}\lambda_{2,\ell}\sum_{\vec{x}}(x_k-\mu)(x_{k+\ell}-\mu)\right];
\end{equation}
i.e., $\log\tilde{f}$ is quadratic in $\vec{x}$, subject to the constraints of unitarity and equations \ref{eq:meanAutocorr}, or 
\begin{equation}
\tilde{f}(\vec{x}) = \left[\left(2\pi\right)^{\dim\vec{x}}\det||{C}||\right]^{-1/2}\exp\left[-\frac{1}{2}\left(\vec{x}-\vec{\mu}\right)^T{C}^{-1}\left(\vec{x}-\vec{\mu}\right)\right]
\end{equation}
where $\vec{\mu}$ is the $\dim\vec{x}$ vector all of whose elements are $\mu$ and the covariance matrix $C^{-1}$ is the related to the autocorrelation function by 
\begin{equation}
{C}_{jk} = c_{k-j}.
\end{equation}
\emph{When nothing more than the noise distributions mean and covariance are known the least presumptive model for the noise statistics is a multivariate Gaussian.} Adopting any other noise model makes additional (and, since we know only $\mu$ and $c_{\ell}$, unjustified) assumptions. This important result is contrary to popular prejudice, which views Gaussian noise models suspiciously even when nothing more is available for use in analysis than the noise mean and covariance. Correspondingly, for the analysis of gravitational wave detector data we have 
\begin{equation} 
\log g(\mathbf{d}|\mathbf{h}',\mathbf{R},\mathcal{N},\mathcal{I})
= 
-\frac{1}{2}\chi^2(\mathbf{d},\mathbf{h}',\mathbf{R},\mathbf{N}) +
\mbox{const.} 
\label{eq:Pchisq}
\end{equation}
where
\begin{equation}
\chi^2(\mathbf{d},\mathbf{h}',\mathbf{R},\mathbf{N})
=
(\mathbf{d}-\mathbf{R}\mathbf{h}')^T \mathbf{N}^{-1}
(\mathbf{d}-\mathbf{R}\mathbf{h}')
\label{eq:chi2}
\end{equation}
may be evaluated directly: i.e., without the need to invert the generally ill-conditioned response $\mathbf{R}$. 

\subsection{The a prior probability $q$}
When the observables $\mathbf{d}$ over-determine the incident wave $\mathbf{h}$ it is tempting to ignore $q(\mathbf{h}'|\mathcal{I})$ in equation \ref{eq:f}, minimize $\chi^2$ (i.e., maximize the likelihood) over $\mathbf{h}'$, and declare that $\mathbf{h}'$ is the inferred incident wave. This is, with minor variation, the approach taken by \citet{gursel:1989:nos}. While an entirely legitimate approach to deconvolution, maximum likelihood methods generally over-fit $\mathbf{d}$: i.e., they find a $\mathbf{h}'$ that leaves a residual $\mathbf{d}-\mathbf{R}\mathbf{h}'$ that is inconsistent with the known noise properties $\mathcal{N}$. One role played by $q(\mathbf{h}'|\mathcal{I})$ is to resist this tendency toward over-fitting noisy data.

One commonly thinks of the a priori probability $q(\mathbf{h}'|\mathcal{I})$ as
a representation of our general expectation signals $\mathbf{h}$ prior to the 
particular observation $\mathbf{d}$. Criticisms of Bayesian analyses generally 
focus on the necessity of invoking a priori probabilities like $q$. A 
great deal of intellectual energy has been devoted to developing ways of 
identifying a priori probabilities that are,  in some meaningful sense, without 
prejudice. The Maximum Entropy Principle, as described above, provides one means
of doing so, which is especially useful when the set of system states is discrete
or the density of states is known. Another approach, also pioneered by \citet{jaynes:1968:pp}, 
takes as its starting point the idea that the prior $q$ should be equivalent 
for equivalent experiments: more formally, that $q$ should be form-invariant under
the same set of transformations that leave the experiment invariant. Such priors
encode only our knowledge of the experiment's nature and do not presuppose 
any outcome. We can use that principle in the present instance to find the class of 
priors $q$ that are the least presumptive regarding the signal $\mathbf{h}$. 

Let us first focus on some general considerations that should govern the prior $q$. Writing $\mathbf{h}$ as the sum of two time series, corresponding to the two polarizations $\mathbf{h}_{+}$ and $\mathbf{h}_{\times}$ of an incident gravitational wave, we note that
\begin{itemize}
\item Lacking any reason to presume that the source is oriented in a
  particular way relative to the detector line-of-sight, the
  prior should be invariant under an arbitrary rotation of $+$ into
  $\times$ polarization;
\item Lacking any reason to presume that the gravitational wave burst
  arrives at a particular moment in time, the prior for $h(t)$ and $h(t+\tau)$ 
  should be the same\footnote{In the case of galactic core-collapse supernova we expect that the arrival time of the observed neutrinos will coincide with the arrival time of the gravitational wave burst up to uncertainties in the time they spend trapped in the dense core. Within these uncertainties we treat the arrival time of the gravitational wave burst as unknown}; and, finally, 
\item Lacking any specific source model that dictates or suggests 
how energy is distributed throughout the burst, the prior should 
not favor waveforms with any particular degree of smoothness (or 
autocorrelation or spectrum). 
\end{itemize}
We could, of course, introduce specific knowledge or assumptions into the construction of the prior, if we have any: for example, we may believe that the gravitational wave power should be concentrated in a band of frequencies, or that there is a relationship between the radiation in the $+$ and $\times$ polarization states, etc. In the present instance, however, our goal is to specify a prior that makes the fewest possible assumptions (beyond propagation direction, which we take to be known) about the nature of the gravitational wave signal incident on our detector(s). 

With these general considerations in mind we may write $\mathbf{h}_{+}$ ($\mathbf{h}_{\times}$) in a Fourier expansion: 
\begin{equation}
{h}_k = \sum_{j} A_j \cos(\omega_j t_k) + B_j\sin(\omega_j t_k).
\end{equation}
The prior on $\mathbf{h}$ can just as well be expressed as a prior $Q$ on the $\vec{A}$ and $\vec{B}$. From our general considerations and the property of Fourier series we conclude that 
\begin{equation}
Q(\vec{A},\vec{B}|\mathcal{I}) = \prod_k \bar{Q}(A_k|\mathcal{I})\bar{Q}(B_k|\mathcal{I}); 
\end{equation}
i.e., the overall prior is the product of priors of identical form evaluated for each of the coefficients $A_k$ and $B_k$. 

\citet{bretthorst:1988:bsa} provides a particularly lucid derivation, which we reproduce here, of the prior $\bar{Q}$ that arises from the transformation group properties that we expect it should satisfy. To begin,
we might just as well have written the Fourier expansion of $\mathbf{h}$ in terms of the Fourier amplitudes $a_j$ and phases $\theta_j$, as in 
\begin{equation}
h_k = \sum_j a_j \cos(\omega_j t_k + \theta_j),
\end{equation}
with the $(a_j,\theta_j)$ related to the $(A_j,B_j)$ by 
\begin{mathletters}
\begin{eqnarray}
a^2_j &=& A_j^2 + B_j^2\\
\tan\theta_j &=& B_j/A_j.
\end{eqnarray}
\end{mathletters}%
The same considerations that led us to factor $Q(\vec{A},\vec{B}|\mathcal{I})$
into the product of identical functions $\bar{Q}$ of the $A_k$ and $B_k$ lead us to factor
the prior $P(\vec{a}, \vec{\theta}|\mathcal{I})$ into the product of priors 
$\bar{P}$ of identical form, evaluated for each of the coefficient pairs 
$(a_k,\theta_k)$:
\begin{equation}
P(\vec{a},\vec{\theta}|\mathcal{I}) = \prod_k\bar{P}(a_k,\theta_k|\mathcal{I})
\end{equation}
Under the assumption that we do not know the signal arrival time all phases $\theta$ are equally likely: i.e.,
\begin{equation}
\bar{P}(a,\theta|\mathcal{I}) = \frac{\bar{p}(a|\mathcal{I})}{2\pi}. 
\end{equation}
The choice of Fourier coordinates $(A_k, B_k)$ or $(a_k,\theta_k)$ does not affect our prior knowledge; so, the prior distributions $\tilde{{P}}$ and $\bar{{Q}}$ must be equivalent: i.e., 
\begin{equation}\label{eq:pq}
\bar{{Q}}(A|\mathcal{I})\bar{Q}(B|\mathcal{I})dA\,dB =
\frac{\bar{{p}}(a|\mathcal{I})}{2\pi} da\,d\theta.
\end{equation}
In the special case where $B$ is equal to zero this relation becomes
\begin{equation}
{\bar{p}(a|\mathcal{I})}\, da\,d\theta = 2\pi\bar{Q}(a|\mathcal{I})\tilde{Q}(0|\mathcal{I})\,dA\,dB. 
\end{equation}
Using this expression for $\bar{p}$ in equation \ref{eq:pq} we find that the desired prior $\bar{Q}$ must satisfy the functional equation
\begin{equation}
\bar{Q}(x|\mathcal{I})
\bar{Q}(y|\mathcal{I}) = \bar{Q}\left(\sqrt{x^2+y^2}\right)\bar{Q}(0|\mathcal{I}),
\end{equation}
which has the general solution 
\begin{equation}
\bar{Q}(x|\sigma^2,\mathcal{I}) = \frac{\exp\left[-x^2/2\sigma^2\right]}{\sqrt{2\pi\sigma^2}}
\end{equation}
for some unknown parameter $\sigma^2$. 

We are thus led to a one parameter family of ``uninformative'' priors $q_\sigma(\mathbf{h}|\mathcal{I})$:
\begin{mathletters}
\begin{eqnarray}
q_\sigma(\mathbf{h}|\mathcal{I}) &=& \prod_k
\tilde{q}_\sigma(A_k|\mathcal{I})
\tilde{q}_\sigma(B_k|\mathcal{I})\\
&=& \frac{\exp\left[-\frac{1}{2\sigma^2}\sum_{k=0}^{N-1}\left(A_k^2+B_k^2\right)\right]}{\left(2\pi\sigma^2\right)^{N/2}}\\
&=& \frac{\exp\left[-\frac{1}{2\sigma^2}\sum_{k=0}^{N-1}h_k^2\right]}{\left(2\pi\sigma^2\right)^{N/2}}. 
\end{eqnarray}
\end{mathletters}%

Setting aside for the moment the question of how we might choose $\sigma$, the choice of prior $q$ taking the form $q_\sigma(\mathbf{h}|\mathcal{I})$ leads us to the posterior probability density 
\begin{eqnarray}
f(\mathbf{h}'|\mathbf{d},\mathbf{R},\mathcal{N},\sigma,\mathcal{I}) &\propto&
g(\mathbf{d}|\mathbf{h}',\mathbf{R},\mathcal{N},\mathcal{I}) q(\mathbf{h}'|\sigma,\mathcal{I})\\
&\propto&\exp\left[-\frac{1}{2}\chi^2(\mathbf{h}',\mathbf{R},\mathbf{N},\mathbf{d}) + \frac{1}{2\sigma^2}S(\mathbf{h}')\right]
\end{eqnarray}
where 
\begin{equation}
S(\mathbf{h}') = -\sum_k {h'}_k^2.
\end{equation}
We may thus choose as the best estimate of $\mathbf{h}$ the $\mathbf{h}'$ that minimizes $\ln f$, or
\begin{equation}
F(\mathbf{h}'|\mathbf{d},\mathbf{R},\mathcal{N},\sigma^2)
= \frac{1}{2}\chi^2(\mathbf{h}',\mathbf{R},\mathbf{N},\mathbf{d}) - \frac{1}{2\sigma^2}S(\mathbf{h}).
\label{eq:functional}
\end{equation}

Stripped of its Bayesian statistical motivation, we recognize that maximum entropy deconvolution of the gravitational signal $\mathbf{h}$ is closely related to the solution of the inverse problem via regularization, with the specific choice $\sum_k h^2_k$ for the regularization function $S$ \citep{neumaier:1998:sia}. Nevertheless, our Bayesian approach to inference is much more than a motivation for a particular choice of regularization function, or even for deconvolution via regularization. Whereas deconvolution provides a single estimate of the signal $\mathbf{h}$, our Bayesian approach provides a distribution $f$, whose mode is the point estimate of the equivalent deconvolution problem but that also assigns a probability to any proposed $\mathbf{h}'$. The availability of this distribution permits, among other things, a detailed error analysis for any problem where the details of the waveform play an important role. 

\subsection{Eliminating the regularization constant}\label{sec:sigma2}
If we view the determination of $\mathbf{h}$ from $\mathbf{d}$ as a problem in deconvolution via regularization, then we require some prescription for the choice of regularization constant $\sigma^2$. Viewed as a problem of Bayesian inference, however, $\sigma^2$ is not unlike any other unknown and we'd prefer, instead, to find some probability distribution $u(\sigma^2)$ that describes the probability that $\sigma^2$ is the appropriate choice and integrate the $q_\sigma$ over $u$ to find a distribution $q$ independent of $\sigma^2$: i.e.,
\begin{eqnarray}\label{eq:marginalize}
q(\mathbf{h}|\mathcal{I})
&=& \int d\sigma^2\, u(\sigma^2) q(\mathbf{h}|\sigma^2,\mathcal{I})
\end{eqnarray}
\emph{Suppose we let the data guide us to a choice of $u(\sigma^2)$:} i.e., we use Bayes' Theorem to construct the probability distribution $u$ of $\sigma^2$, based on the observations $\mathbf{d}$:
\begin{equation}
u(\sigma^2|\mathbf{d},\mathcal{N},\mathbf{R},\mathcal{I}) \propto
\int D\mathbf{h}\,
g(\mathbf{d}|\mathbf{h}',\mathbf{R},\mathcal{N},\mathcal{I}) 
q(\mathbf{h}'|\sigma^2,\mathcal{I})
t(\sigma^2|\mathcal{I})
\end{equation}
where we have introduced the prior $t(\sigma^2|\mathcal{I})$.

In general we have no knowledge that suggests a preferred value of $\sigma^2$; correspondingly, any reasonable prior $t$ should be approximately constant over a wide range of $\sigma^2$. Additionally, when the data $\mathbf{d}$ are informative then $gq_\sigma$ will be a sharply peaked (relative to $t$) function of $\sigma^2$.  For both these reasons we may treat $t$ as constant over the support of $gq_\sigma$ as a function of $\sigma^2$ in the integrand of equation \ref{eq:marginalize}, in which case
\begin{equation}
u(\sigma^2|\mathbf{d},\mathcal{N},\mathbf{R},\mathcal{I}) \propto
\int D\mathbf{h}\,
g(\mathbf{d}|\mathbf{h}',\mathbf{R},\mathcal{N},\mathcal{I}) 
q_\sigma(\mathbf{h}'|\mathcal{I})
\end{equation}
This integral we recognize as the normalization constant $v$ that appears in our expression for
$f$ (cf.~eq.~\ref{eq:f}) when evaluated for a particular value of $\sigma$. \emph{In addition to allowing us to infer the signal $\mathbf{h}$ our observations $\mathbf{d}$ also provide us with a probability distribution for the ``regularization constant'' $\sigma^2$.}

For our problem the distribution $u(\sigma^2|\mathbf{d},\mathcal{N},\mathbf{R},\mathcal{I})$ involves only Gaussian integrals and is relatively straightforward to calculate: recalling our definition of $\chi^2$ (cf.\ eq.\ \ref{eq:chi2}) and introducing 
\begin{mathletters}
\begin{eqnarray}
Z_g &=& \int D\mathbf{d}\exp \left(-\frac{1}{2}\chi^{2}\right) = 
\left[\left(2\pi\right)^{N_d}\det||\mathbf{N}||\right]^{1/2}\\
Z_q &=& \int D\mathbf{h} \exp\left(-\frac{\mathbf{h}^2}{2\sigma^2}\right) = 
\left(2\pi\sigma^2\right)^{N_h/2}\\
Z_f &=& \int D\mathbf{h} \exp\left(-\frac{1}{2}\chi^{2}-\frac{\mathbf{h}^2}{2\sigma^2}\right) = 
\left[
\frac{\left(2\pi\right)^{N_h}}{\det||\sigma^{-2} \mathbf{I} + \mathbf{R}^T\mathbf{N}^{-1}\mathbf{R}||}
\right]^{1/2}
\end{eqnarray}
we find
\begin{eqnarray}
2\log u &=& -\chi^2(\mathbf{d},\mathbf{R}\mathbf{h}_0,\mathcal{N}) -\sigma^{-2}S(\mathbf{h}_0) 
+ 2\log\frac{Z_f}{Z_{g} Z_{q}}\nonumber\\
&=& 
-\chi_0^2 -\frac{S_0}{\sigma^{2}}
-N_d\log2\pi 
+2N_h\log\sigma
-\log\det||\sigma^{-2}\mathbf{I} + \mathbf{R}^T\mathbf{N}^{-1}\mathbf{R}||,
\label{eq:logDet}
\end{eqnarray}
where $\mathbf{h}_0$ is the $\mathbf{h}$ that maximizes $g(\mathbf{h})q_\sigma(\mathbf{h})$, $N_d$ is $\dim\mathbf{d}$, and $N_h$ is $\dim\mathbf{h}$
\end{mathletters}%

We can generally approximate equation \ref{eq:marginalize} for $q(\mathbf{h}|\mathcal{I})$ by $q_{\hat{\sigma}}(\mathbf{h}|\mathcal{I})$, where $\hat{\sigma}^2$ maximizes $u(\sigma^2|\mathcal{I})$. While conceptually we do not make a choice of $\sigma$, as a matter of practice if a signal is recoverable $f$ depends at most weakly on $q$ and the desired prior $q$ is also generally close to $q_{\hat{\sigma}}$. To find $\hat\sigma$ we can extremize our expression for $\log u$ (cf. \ref{eq:marginalize}) over $\sigma$:
\begin{mathletters}
\begin{equation}
0 = \frac{d\log u}{d\sigma} = 
\frac{S_0}{\sigma^3} - \frac{N_h}{\sigma} + \sigma^{-3}\mathrm{tr}\left[\left(\sigma^{-2}\mathbf{I}+\mathbf{R}^T\mathbf{N}^{-1}\mathbf{R}\right)^{-1}\right]
\end{equation}
or 
\begin{equation}\label{eq:hatSigma}
\hat\sigma^2= 
\frac{\sum_{k} h_k^2}{
N_h - 
\mathrm{tr}\left[\left(\mathbf{I}+\hat\sigma^{2}\mathbf{R}^T\mathbf{N}^{-1}\mathbf{R}\right)^{-1}\right]
},
\end{equation}
where we made use of the relation
\begin{equation}
\ln\det ||\mathbf{A}|| = \mathrm{tr}\ln||\mathbf{A}||.
\end{equation}
\end{mathletters}%
Equation \ref{eq:hatSigma} provides an implicit relationship for $\hat{\sigma}^2$, which can be solved iteratively with the 
minimization of equation \ref{eq:functional} for $\mathbf{h}$: e.g., we can solve 
equation \ref{eq:functional} for $\mathbf{h}$ given a guess for $\hat{\sigma}^2$ 
and use this $(\hat{\sigma}^2,\mathbf{h})$ pair on the right-hand side of equation 
\ref{eq:hatSigma} to find a new estimate for $\hat{\sigma}^2$, with the process 
repeated until it converges. 

\subsection{Example}\label{sec:Example}
As an example, consider a plane gravitational wave signal propagating from a hypothetical source directly overhead of the LIGO Hanford Observatory (LHO) site, and incident on detectors at the LIGO Hanford and LIGO Livingston observatory sites. For this example assume that 
\begin{itemize}
\item Each site has a single, identical interferometric gravitational wave detector oriented in the same way as the actual detectors at each site; 
\item The response function for each detector is frequency independent in the band of interest; 
\item The detector noise is independent between detectors and white (i.e., the autocorrelation is a delta function in lag), with unit variance in each detector;
\item The sample rate at each detector is 4096~Hz.
\end{itemize}
For the signal use
\begin{mathletters}
\begin{eqnarray}
\mathbf{h} &=& h_{+}\mathbf{e}_{+} + h_{\times}\mathbf{e}_{\times}\\
h_{+} &=& h_0 \cos\phi_0 e^{-(t-t_0)^2/2\sigma_t^2}\cos[2\pi f (t-t_0)]\\
h_{\times} &=& h_0 \sin\phi_0 e^{-(t-t_0)^2/2\sigma_t^2}\sin[2\pi f(t-t_0)]\\
\left|t-t_0\right| &<& 24.4\,\mbox{ms}\\
f &=& 414\,\textrm{Hz}\\
\sigma_t &=& 4\,\textrm{ms}
\end{eqnarray}
\end{mathletters}%
with gravitational wave polarization tensor $\mathbf{e}_{\times}$ orthogonal to tangents to lines of latitude, $\mathbf{e}_{+}$ orthogonal to $\mathbf{e}_{\times}$, and $\phi_0$ is chosen so signal projection on the LHO detector is maximized. 

\begin{figure}
\plotone{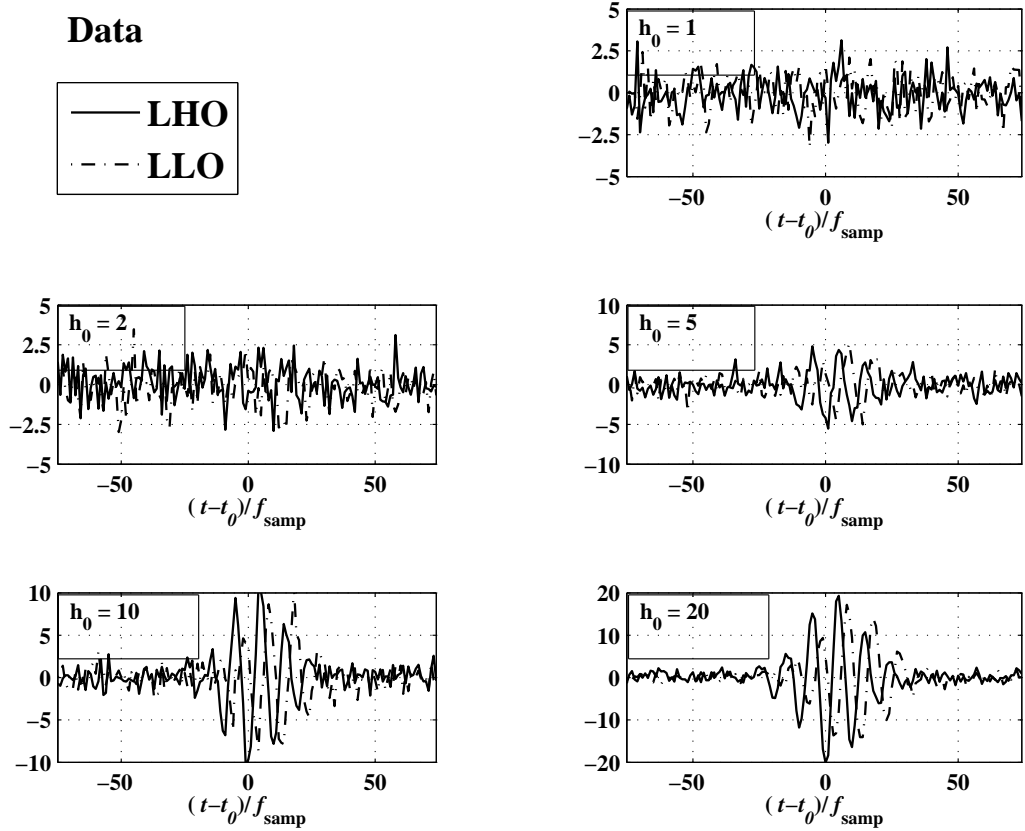}
\caption{A short snippet of the simulated data surrounding the embedded signal in the simulations described in \S\ref{sec:Example}. The embedded signal was approximately 100 samples long and centered in the snippet shown. The demonstration analysis was carried out on a data segment five times longer than the signal. See \S\ref{sec:Example} for more details.}
\label{fig:exampleD}
\end{figure}

\begin{figure}
\plotone{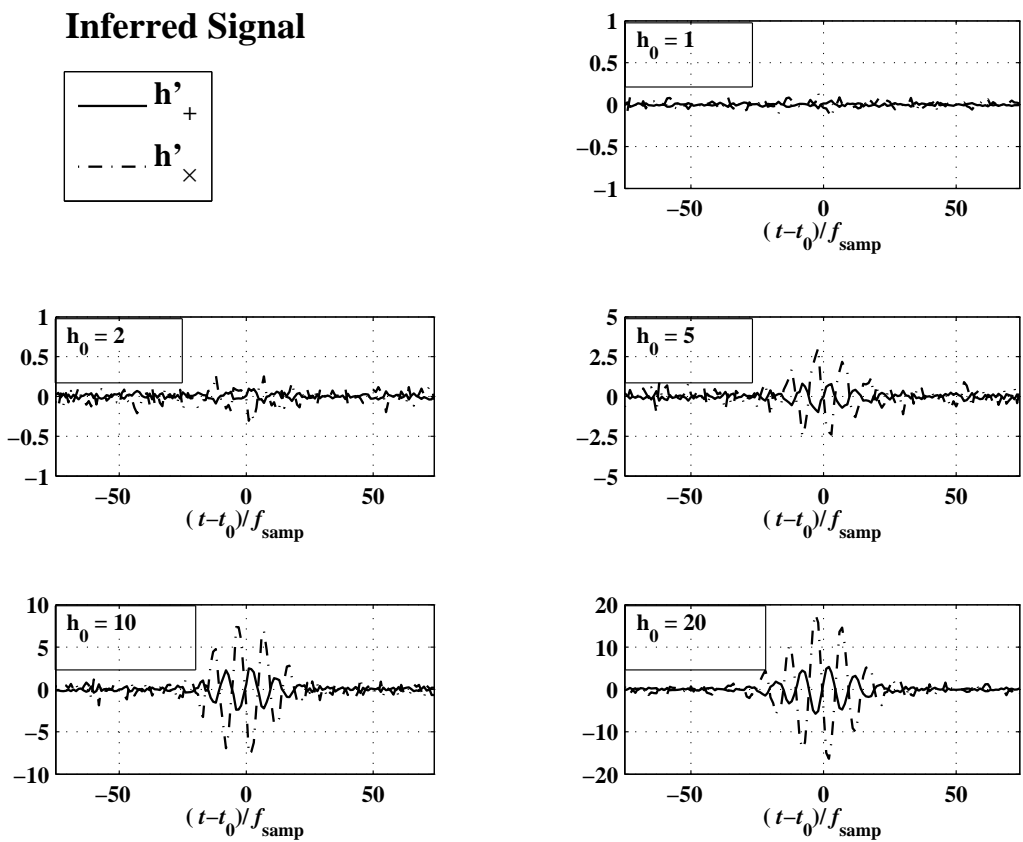}
\caption{The inferred gravitational wave signal in the $+$ and $\times$ polarization states for the five data sets described in \S\ref{sec:Example}. Each panel here corresponds to the data shown in the corresponding panel of figure \ref{fig:exampleD}. Note that the two detector network does an excellent job of rejecting detector noise (which has unity variance) and that the actual signal emerges very rapidly once the signal amplitude begins to exceed the noise amplitude. See \S\ref{sec:Example} for more details.}
\label{fig:exampleH}
\end{figure}

\begin{figure}
\plotone{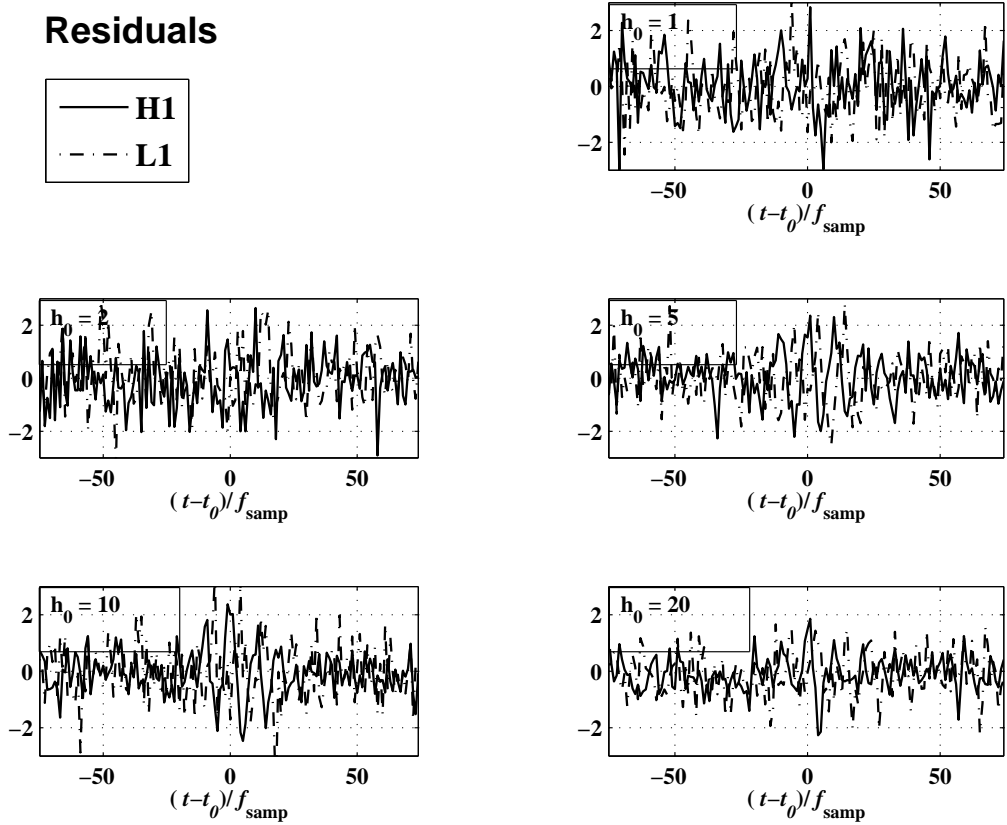}
\caption{The residuals $\mathbf{d}-\mathbf{Rh}'$ following subtraction of the response of the simulated detector network to the inferred signal ($\mathbf{Rh}'$) from the observations ($\mathbf{d}$) made in the network, as described in \S\ref{sec:Example}.  Each panel here corresponds to the data shown in the corresponding panel of figures \ref{fig:exampleD} and \ref{fig:exampleH}. The amplitude of the residuals is consistent with detector noise and the temporal structure of the residuals is, in the absence of additional knowledge regarding either the actual signal or the detector noise, also consistent with the detector noise. See \S\ref{sec:Example} for more details.}
\label{fig:exampleResid}
\end{figure}

Figure \ref{fig:exampleD} shows snippets of the simulated data from five different data sets $\mathbf{d}$. Each data set corresponds to the (noisy) observation at the detector network described above when a signal of amplitude $h_0$ equal to 1, 2, 5, 10, or 20, is incident on the detector network. Each snippet is centered on the location of the actual signal, which was 100 samples in duration and embedded in a data set five times as long. 
Figure \ref{fig:exampleH} shows the most probable waveforms $h'_{+}$ and $h'_{\times}$ as inferred, in the manner just described, from each data set. Finally, figure \ref{fig:exampleResid} shows the residual $\mathbf{d}-\mathbf{Rh}'$ for each data set and inferred waveform $\mathbf{h}'$. Table \ref{tbl:example} shows the power signal-to-noise ratio $\rho^2$ of the recovered signals, calculated as\footnote{It is important to note that the signal-to-noise as defined here involves only the inferred signal: i.e., it is the observed signal-to-noise, which varies from simulation to simulation with the different instantiations of the noise. Calculated in this way it is a different quantity than the signal-to-noise often quoted in the context of experimental results published by the LIGO Scientific Collaboration, who report their sensitivity in terms of the \emph{expectation value} of the signal-to-noise for a \emph{known} signal --- not an inferred one --- supposed to be present in the data (see, for example \citet{abbott:2004:aol}).}
\begin{equation}\label{eq:snr}
\rho^2 = \left(\mathbf{Rh}'\right)^{T}\mathbf{N}^{-1}\left(\mathbf{Rh}'\right) = \chi^2(0,\mathbf{h}',\mathbf{R},\mathbf{N})
\end{equation}
and the correlation between the actual and the inferred signal, calculated as
\begin{mathletters}\label{eq:xcorr}
\begin{equation}
C(j) =
\frac{\sum_k{h_{k+j}h'_{k}}}%
{\left|\mathbf{h}'\right|\,\left|\mathbf{h}\right|} 
\end{equation}
where
\begin{equation}
\left|\mathbf{x}\right|^2 = \sum_{k}x_k^2.
\end{equation}
\end{mathletters}%
The value of $\rho^2$ and $C(j)$ recorded here should be taken as representative: their actual values depend on the noise instantiation, which varies from simulation to simulation (and from observation to observation). 

Comparing figures \ref{fig:exampleD} and \ref{fig:exampleH} we see the degree to which out inference procedure is able to separate signal from noise. For the cases $h_0$ equal to 1 or 2 no signal is resolved: i.e., $h_0$ is uniformly small and, in fact, much smaller than the rms noise. For the three cases $h_0$ equal to 5, 10 and 20 the inference procedure clearly does resolve distinct signals in each of the two gravitational wave polarization. The maximum cross-correlations (cf. equation \ref{eq:xcorr}), tabulated in table \ref{tbl:example}, makes the correctness of the inference as a function of $h_0$ quantitative: the correlation between the real and inferred signals is greater than a half for signal amplitudes greater than 5, and inspection of figure \ref{fig:exampleH} shows that the error is principally in the signal amplitude and not its time dependence. 

Should we expect to be able to do better? Estimates of the required power signal-to-noise for reliable detection of a gravitational wave burst in LIGO based solely on the coincidence of excess power in the detector data streams\footnote{And calculated in a manner comparable to the $\rho^2$ recorded in table 1} range from 30 to 70 \citep{thorne:1987:gr,finn:1991:dog,finn:1992:dma}. Referring to table \ref{tbl:example} it is apparent that a reliable signal emerges very rapidly between signal amplitudes $h_0$ of 2 and 5, corresponding to $\rho^2$ in this range. Figure \ref{fig:exampleResid} shows $\mathbf{Rh}'-\mathbf{d}$, the residual after the inferred signal is subtracted from the observations. If we have done a good job of separating the signal and noise contributions to the observations $\mathbf{d}$ than we expect that $\mathbf{Rh}'-\mathbf{d}$ should be noise: in this example, unity rms white noise. In the cases $h_0$ equal to 1 or 2 inspection of the residuals shows no obvious evidence of structure left behind by the inference procedure. For the case $h_0$ equal to five and greater, where a clear signal is identified, the residual does show some evidence of structure near the location of the signal peak. The amplitude of this apparent structure is very much consistent with our assumptions about the detector noise (i.e., unity rms): it is only the time dependence of the structure that draws the attention of the eye. Over longer random time series, even this would not appear to be particularly unusual: i.e., were we not aware that we were looking at a residual we would not have reason based on statistics to be particularly suspicious of the presence of a signal here. Reviewing the construction of our inference procedure, the quantitative agreement between our estimates of the required signal-to-noise for a reliable detection, and the qualitative evidence of these residuals, it is clear that absent additional information about the nature of the noise (which would modify our function $\chi^2$) or the nature of the signal (which would modify our choice of function $S$) we cannot reasonably suppose that there is any further information in the residuals about a possible signal. 

\subsection{Summary}
The problem of identifying a signal in noisy data can be approached as a problem of statistical inferrence: i.e., we can find a probability distribution that describes the credibility that we should ascribe to the hypothesis that an arbitrary gravitational wave burst $\mathbf{h}$ was incident on an array of detectors given observations $\mathbf{d}$. This probability distribution involves two components: the probability $g$ that $\mathbf{d}-\mathbf{R}\mathbf{h}$ is detector noise, and
the a priori probability $q$ that a signal takes the form $\mathbf{h}$. 

The probability $g$, also known as the likelihood function, requires a characterization of the noise. In the absence of a complete characterization of the noise --- something that is rarely, if ever, possible --- the Maximum Entropy Principle provides a means of selecting the probability $g$ that is simultaneously consistent with what is known about the noise and least presumptive about what is not. In particular, when the noise is characterized by its mean and covariance --- the usual case --- then $g$ takes the form of a Normal distribution. 

The a priori probability density $q$ characterizes our prior expectations regarding signals $\mathbf{h}$. In the spirit of ``letting the data speak for itself'' we avoid prejudicing our analysis with preconceived notions regarding the nature of potential sources by focusing instead on our ignorance of signal polarization, arrival time, and energy spectrum. In this way we are led to a one-parameter family of potential priors $q_\sigma$. Rather than choose a particular prior from this family, we find that the data allow us to provide probabalistic weights $u(\sigma)$ to different values of the parameter $\sigma$. Rather than make a particular choice of distribution $q_\sigma$, we evaluate the expectation value of $q_\sigma$ over the distribution of the parameter $\sigma$ and use it for $q$, eliminating our dependence on the unknown parameter $\sigma$. As a practical matter, it is often sufficient to find the $\hat\sigma$ that maximizes $u(\sigma)$ and use $q_{\hat\sigma}$ for $q$.

\section{APPLICATION TO LIGO OBSERVATIONS OF SN}
\label{sec:Application}

How well can we hope to characterize the astrophysics of core-collapse
supernovae from the gravitational wave signature we observe?  The
detailed astrophysics of core-collapse supernovae is uncertain,
difficult to model, and involves large-scale convection and other
stochastic processes. For all these reasons, it is more likely that we
will infer the features of the gravitational wave burst associated with a
supernova and then use that inferred signal to validate our models, than
that we will ``discover'' the gravitational wave signal from a
core-collapse supernova by using matched filtering to extract it from
deep within a noisy data stream.

As a first step toward exploring how, and how well, we may be able to
diagnose the conditions of the collapsing core from the gravitational
waves it radiates we use the techniques, described in the previous two
sections, to infer a simulated signal embedded in LIGO-like noise and
evaluate the cross-correlation between the inferred signal and a wide
range of signals calculated from the models of \citet{ott:2004:gwf}.
We do this for
\begin{itemize}
\item LIGO data drawn from the S1--S4 science runs, evaluating the
  maximum distance at which there is a significant correlation between
  the inferred and simulated signals (cf. sec. \ref{sec:LIGO});
\item Simulated signals with varying maximum central density (cf. sec.
  \ref{sec:density}), evaluating our ability to evaluate the
  properties of the core-collapse;
\item Simulated signals with varying progenitor mass, evaluating our
  ability to test the correlation between progenitor models and the
  configuration of the collapsing stellar core (cf. sec.
  \ref{sec:mass});
\item Simulated signals with varying angular momentum and differential
  angular velocity, allowing us to explore the role that differential
  rotation plays in core collapse (cf. sec. \ref{sec:Omega}).
\end{itemize}
We find that, even when the simulated signal is so weak that the
inferred signal appears very different, the inferred signal still has
its greatest cross-correlation with the simulated signal. We conclude
that the model whose signal has the maximum cross-correlation with the
inferred signal is likely to provide a good indication of the physical
properties of the source.

\subsection{The Gravitational Wave Signature of Core-Collapse Supernovae:
Description of Core-Collapse Models}
\label{sec:Ott}

Gravitational wave emission from core-collapse supernovae
may arise from a multitude of processes, including  
rotating core collapse and core bounce
(e.g., \citealt{fryer:2002:gwe,dimmelmeier:2002:rso,ott:2004:gwf,dimmelmeier:2007:ggs,dimmelmeier:2007:ggs:1,ott:2007:3co}), 
postbounce convective overturn, 
anisotropic neutrino emission (\citealt{burrows:1996:pra,muller:2004:tgw,ott:2007:3co}), 
nonaxisymmetric rotational instabilities of the protoneutron star
(\citealt{rampp:1998:son,ott:2007:rco,shibata:2005:tso}), or from the
recently proposed protoneutron star core g-mode oscillations
(\citealt{burrows:2006:nmf,burrows:2007:foa,ott:2006:nmf,ferrari:2003:gwf}).
In addition and in the context of the core-collapse supernova
$-$ gamma-ray burst connection \citep{woosley:2006:sbc}, late-time
black hole formation in a failed or weak core-collapse
supernova may accompany gravitational wave emission from
quasi-normal ring-down modes of the newly-formed black hole.

Of all the above emission processes, rotating iron core collapse
and bounce is the most extensively modeled and best quantitatively and
qualitatively understood. For these reasons we limit our present
study to an analysis of the rotating core collapse and bounce
signature only and use example templates computed from a set of 
2D axisymmetric Newtonian core collapse simulations 
by~\citet{ott:2004:gwf} which focused on the dynamics of 
rotational collapse and bounce.  These represent a class of
signatures of core collapse and are good templates with 
which to exercise the signal extraction technology we
have developed.  This paper is the first to put theoretical
models of the gravitational wave signals of core collapse and bounce 
through a realistic detector pipeline and to attempt to extract
physical information using sophisticated signal processing
algorithms.

The astrophysics models involved stellar progenitors with various masses: 11,
15, 20, and 25$M_\odot$ calculated in~\citet{woosley:1995:eae}. The
simulations neglected the effects of neutrinos, general relativity,
and magnetic fields, but used the realistic, finite-temperature 
nuclear equation of state
of~\citet{lattimer:1991:geo}.  A small number of simulations also
investigated stellar progenitor models from~\citet{heger:2000:peo}
and~\citet{heger:2004:peo}, which were evolved to the onset of iron
core collapse with an approximate treatment of
rotation~\citep{heger:2000:peo,heger:2004:peo} and angular momentum
redistribution by magnetic torques~\citep{heger:2004:peo}.  The
gravitational wave signature extraction was performed using the
Newtonian quadrupole formalism (see e.g.~\citealt{misner:1973:g}).
 
The effects of rotation were investigated 
in~\citet{ott:2004:gwf}. The initial rotation of the progenitor was
controlled by two parameters: the rotation parameter $\beta$ where
\begin{equation}
\label{eq:beta}
\beta = \frac{E_{\mathrm{rot}}}{\mid E_{\mathrm{grav}} \mid} ,
\end{equation} 
and the differential rotation scale parameter $A$, which is the
distance from the rotational axis at which the rotational velocity
drops to half that at the center.  $A$ is defined as
\begin{equation}
\label{eq:A}
\Omega(r) = \Omega_0 \left[ 1 + \left(\frac{r}{A}\right)^2 \right]^{-1} ,
\end{equation}
where $r$ is the distance from the axis of rotation and $\Omega(r)$ is
the angular frequency at $r$.

When the progenitor is rotating slowly and $\beta$ is small (zero to a
few tenths of a percent), the collapse is halted when the inner core
reaches supranuclear densities. The core bounces rapidly and then
quickly rings down.  When the progenitor rotates more rapidly and
$\beta$ is larger, the core collapse is halted by centrifugal forces
and the core bounces at subnuclear densities. The core then undergoes
multiple damped, harmonic oscillator-like expansion-collapse-bounce
cycles\footnote{Recent results
of \citet{ott:2007:rco} and \citet{dimmelmeier:2007:ggs,dimmelmeier:2007:ggs:1} suggest that such
multiple-bounce dynamics are less likely when general
relativity and deleptonization are taken into account.}
The initial degree of differential rotation affects the value
of $\beta$ at which this bounce type transition occurs. A progenitor
with a smaller value of $A$ experiences a greater amount of
differential rotation and hence, a more rapidly rotating inner core.
As a result, the transition from a supranuclear to a subnuclear bounce
occurs for a lower value of $\beta$.

The models of \citet{ott:2004:gwf} yield absolute values of the 
dimensionless maximum gravitational wave strain $h_\mathrm{max}$ 
in the interval 2$\times$10$^{-23}$ $\le h_\mathrm{max} \le$
1.25$\times$10$^{-20}$ at a detector distance of 10 kpc. The
total energy radiated ($E_\mathrm{GW}$) lies in the range
1.4$\times 10^{-11}\, \mathrm{M}_\odot c^2$ $\le E_\mathrm{GW} \le$ 
2.21$\times$10$^{-8}\, \mathrm{M}_\odot c^2$ and most of it
is emitted in the primary gravitational wave burst associated
with core bounce.  The energy spectra
peak in the frequency interval 20~Hz $\le f_\mathrm{peak} \le$
600~Hz with rapid and differential rotators having peaks at low
frequencies and moderate and rigid rotators peaking at high
frequencies.

\subsection{Gravitational wave observations of rotating 
stellar core collapse}
\label{sec:simulation}

In all the simulations described here we used \citet{ott:2004:gwf} model 
s15A1000B0.1 as the ``real'' signal, and correlated the inferred signal with 
other models in the \citet{ott:2004:gwf} catalog. Model s15A1000B0.1 
corresponds to a 15~$\mathrm{M}_{\odot}$~\citet{woosley:1995:eae} 
progenitor with rotation parameter $\beta$ equal to 0.1\% and differential 
rotation scale parameter $A$ equal to 1000~km. We scaled this signal to 
represent core collapse events at different distances and projected the 
incident signal onto the LIGO 4-km Hanford WA (LHO) and Livingston, LA 
(LLO) detectors. For the purpose of this study we assumed that the core 
collapse was directly overhead of the LHO site.\footnote{In this work we have always
assumed that the gravitational wave signal we observe is accompanied by
a neutrino or electromagnetic signal that determines the source sky position.} 
Since the~\citet{ott:2004:gwf} 
core collapse models are axisymmetric the gravitational waves 
they emit are linearly polarized. We chose the polarization angle to 
maximize the response of the LHO detector and used the actual 
response functions for the LHO and LLO detectors characteristic of 
LIGO's S4 science run \citep{gonzalez:2004:col}. We simulated the 
detector noise by adding white noise with power spectral density 
amplitude approximately equal to the noise amplitude at 100~Hz in the 
corresponding science run~\citep{lazzarini:2005:sss}.\footnote{At the 
frequencies where the signal power (in units of squared strain) of the 
simulated core collapse event gravitational wave signals peaks ($\sim$500 
Hz) the noise power spectral density in any of the current generation of 
interferometric gravitational wave detectors is increasing relatively slowly 
with frequency. (Very little is known concerning the cross-correlation in the 
noise of two separated detectors, other than that it is very small.) While we 
have chosen to make this demonstration with white noise, the maximum 
entropy analysis method introduced above naturally accommodates any 
noise covariance: cf. the paragraph including equation (\ref{eq:Pchisq}).} 
Finally, we used the maximum entropy method described above to find the 
best point estimate of the embedded signal and cross-correlated this 
estimate signal with different signals drawn from the~\citet{ott:2004:gwf,ott:2007:3co} 
parameter survey. The subsections below describe our observations based 
on this study.

\subsubsection{Science Run and Survey Range}\label{sec:LIGO}

Before discussing how well we can distinguish between different core
collapse models we looked at how close a core collapse event 
would have needed to be for LIGO, during its first four
science runs (S1, S2, S3 and S4), to infer its waveform with reasonable
accuracy. For this purpose we used the detector response
functions and noise power spectral density amplitudes for the corresponding science run~\citep{%
adhikari:2003:col,%
gonzalez:2004:col,%
lazzarini:2002:lss,%
lazzarini:2003:lss,%
lazzarini:2004:sbs,%
lazzarini:2005:sss%
}.  

\begin{figure}
  \plotone{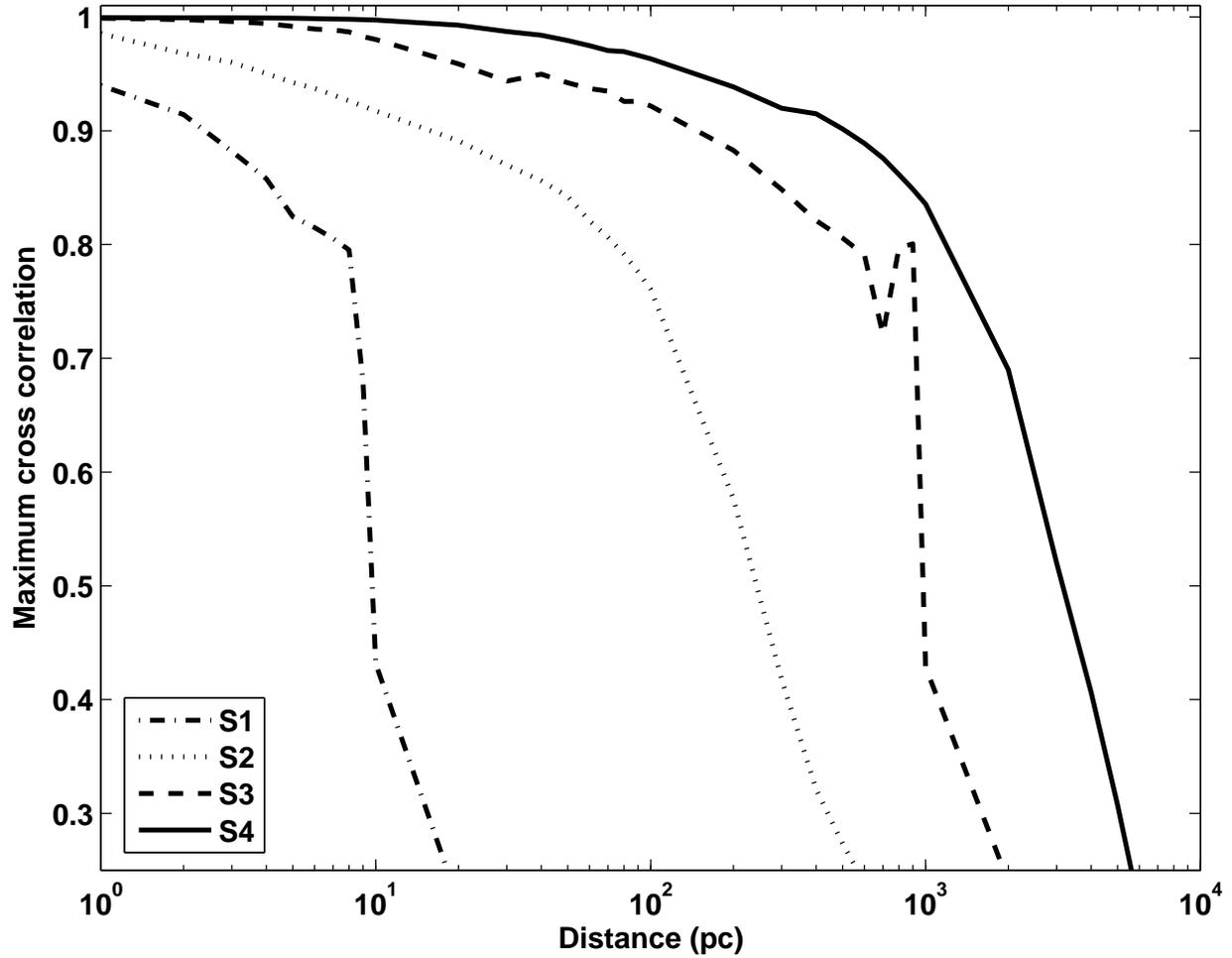}
\caption{Maximum cross-correlation between estimated and initial
  signals versus core collapse distance for data simulated using different
  science run detector impulse responses and noise levels.  The
  estimated signal is inferred using maximum entropy from simulated
  detections that use~\citet{ott:2004:gwf} model s15A1000B0.1 as the
  initial signal waveform.  The up-tick in the S3 run, located at
  approximately 1~kpc, is the result of a discontinuous change in the
  most probable waveform. There is a steady improvement in maximum
  entropy's ability to reconstruct fainter, more distant signals as
  the sensitivity of the detectors improved~\citep{%
lazzarini:2002:lss,%
lazzarini:2003:lss,%
lazzarini:2004:sbs,%
lazzarini:2005:sss}.}\label{fig:scirun}
\end{figure}

Figure~\ref{fig:scirun} shows the maximum cross-correlation, for each
of the first four science runs, between the inferred and actual
waveforms as a function of the core collapse distance. There is a steady
improvement, from S1 to S4, in maximum entropy's ability to recover
signals at greater distances, corresponding to improving detector
sensitivity: by S4, we are able to infer the gravitational waveform
from core collapse events 
that occur as far as a few kpc away. The up-tick in
cross-correlation at the 1~kpc mark in curve for S3 sensitivity is the
result of a discontinuous change in the most probable waveform (i.e.,
a new extrema becoming the global maximum in the probability
function.) The investigations described in the following sections, which delve
into the source information present in the inferred signal, use the
response and noise characteristic of the LIGO S4 science run. 

An alternative way of expressing the results summarized in Figure~\ref{fig:scirun} is to describe the sensitivity required of a LIGO-like detector to observe an optimally located supernovae to a given distance. Under the ``white noise'' approximation made here, doubling the distance to which LIGO is able to infer the waveform from a core collapse supernova requires doubling its sensitivity: i.e., halving the rms noise amplitude at 100~Hz. Referenced to LIGO's design sensitivity at 100~Hz, which was achieved during LIGO's S5 science run, and a 70\% maximum cross-correlation, the maximum distance at which core collapse supernova are expected to be observable is 
\begin{equation}
d_{\mathrm{max}}(70\%) = 4\,\mathrm{kpc}\left(\frac{3\times10^{-23}\,\mathrm{Hz}^{-1/2}}{\sqrt{S_h(100\,\mathrm{Hz})}}\right)
\end{equation}
where $S_h(f)$ is the strain-equivalent noise power spectral density. 
The intermediate LIGO upgrade underway at this writing is expected to reduce $S_h$ by a factor of four, and the advanced LIGO upgrade that will follow will reduce $S_h$ by another factor of approximately 100 \citep{ligo:2007:web}. Correspondingly, we expect that advanced LIGO will be capable of observing core-collapse supernovae like those modeled here at distances as great as 80~kpc: i.e., beyond the distance to the LMC and SMC, but far short of the Virgo cluster. 

\subsubsection{Bounce Type}\label{sec:density}

\begin{figure}
  \plotone{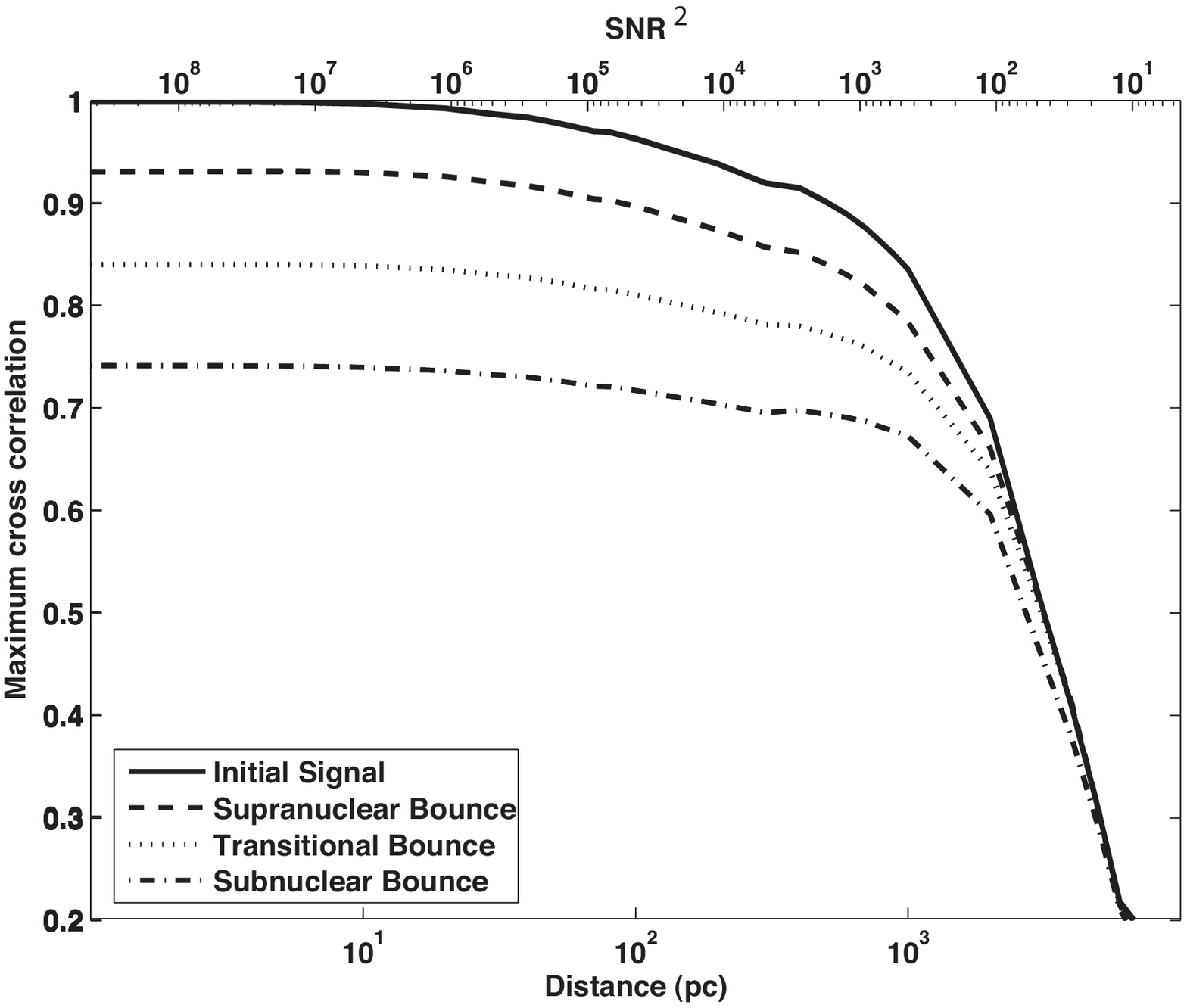}
\caption{Maximum cross-correlation between reconstructed
  waveforms and waveforms associated with models that differ by bounce
  type versus core collapse distance and $\textrm{SNR}^2$. The $\textrm{SNR}^2$ shown is
  the average for the two detectors.  The $\textrm{SNR}^2$ for the 4-km
  Hanford detector is 1.05 times that shown while the $\textrm{SNR}^2$ for the
  4-km Livingston detector is 0.95 times that shown.  The
  reconstructed waveform is inferred using maximum entropy from
  simulated detections that use the waveform from
  the~\citet{ott:2004:gwf} model s15A1000B0.1 as the initial signal
  waveform as well as detector responses and noise levels from the
  fourth science run (S4). The solid line represents the maximum
  cross-correlation between the reconstructed signal and the initial
  signal waveform.  The other lines represent the maximum
  cross-correlations between the inferred waveforms and the waveforms
  resulting from each bounce type for which the maximum
  cross-correlation at 1 pc is greatest, excluding that used for the
  initial signal.  The inferred waveform is most similar to those
  generated by models with the same, supranuclear bounce type as the
  initial signal waveform, for the simulations corresponding to
  core collapse events that occur less than 2-3 kpc away.}
\label{fig:bouncetype}
\end{figure}

\begin{figure}
  \plotone{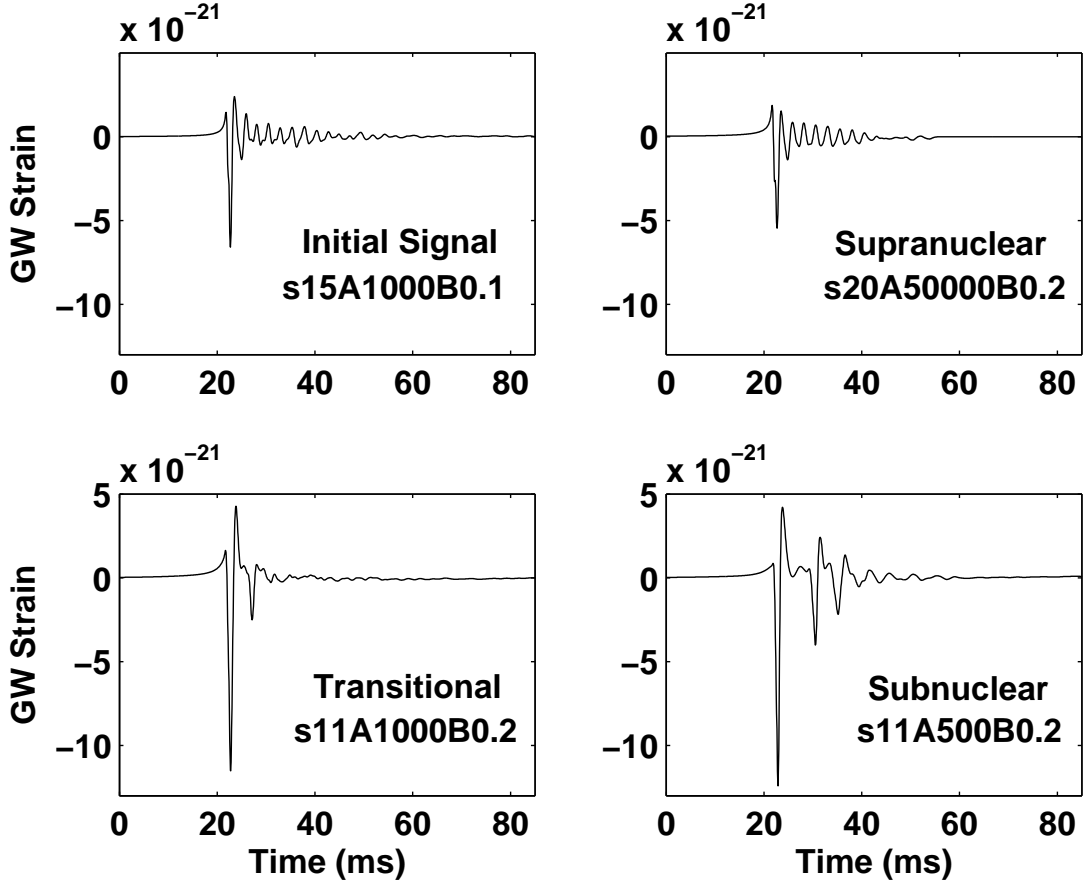}
 \caption{The waveforms associated with various bounce types that
   are compared with the inferred signal in Fig.~\ref{fig:bouncetype}.
   The upper left plot shows the waveform (from Ott et al. [2004]
   model s15A1000B0.1) that was used as the initial signal in the
   simulated detection.  The three other waveforms shown are those
   that are most similar to this initial signal waveform, for each
   bounce type.  The waveform from the~\citet{ott:2004:gwf}
   s20A50000B0.2 model looks much like the initial signal waveform
   which is of the same, supranuclear bounce type.  The subnuclear
   bounce waveform shows the effects of multiple damped, harmonic
   oscillator-like expansion-collapse-bounce cycles. The zero points
   of the time axes are chosen so that the minima of the waveforms
   occur at the same time for ease of comparison.  The waveform
   amplitudes are scaled to correspond to core collapse events
   at 10 kpc.}
        \label{fig:wave_bouncetype}
\end{figure}

We can classify the models of~\citet{ott:2004:gwf} into those that
bounce at supranuclear, subnuclear and transitional central densities.
Figure~\ref{fig:bouncetype} shows the maximum cross-correlation
between the inferred waveform, the actual waveform (a supranuclear
bounce type), and three other waveforms that each have, within their
respective categories (supernuclear, subnuclear and transitional), the
greatest cross-correlation with the inferred waveform. (Figure~\ref{fig:wave_bouncetype} shows these four waveforms
themselves, each with strain scaled to a distance of 10 kpc.)
It is clear 
that the inferred waveform has the most in common with that generated 
from a model with the same, supranuclear, bounce type and that, for 
S4 detector sensitivities, our ability to distinguish bounce type fails for
core collapse events more than 3~kpc distant.

To facilitate comparison with signal-to-noise ratio sensitivities as reported by the LIGO Scientific Collaboration for the LIGO detectors, we show the quantity ``$\textrm{SNR}^2$'', 
defined by
\begin{mathletters}
\begin{eqnarray}
\mathrm{SNR}^2 &=& \frac{1}{2}\left(H_1^2 + H_2^2\right),\\
H_k^2
&=&\sum_{j=0}^N\frac{\left(\mathbf{r}_k\mathbf{h}\right)^2}{N\sigma_k^2}, 
\end{eqnarray}
\end{mathletters}%
where $h$ is the \emph{actual} waveform, 
running across the top of this and subsequent figures. The quantity $\textrm{SNR}^2$ can be compared to the single-detector, optimal orientation, expectation value of the power signal-to-noise ratio sensitivities reported by LIGO in its publications: i.e., assuming that the actual gravitational waveform were known, the expectation value of the signal-to-noise determined by matched filtering would be given by $\textrm{SNR}$. 

\subsubsection{Mass}\label{sec:mass}
\begin{figure}
  \plotone{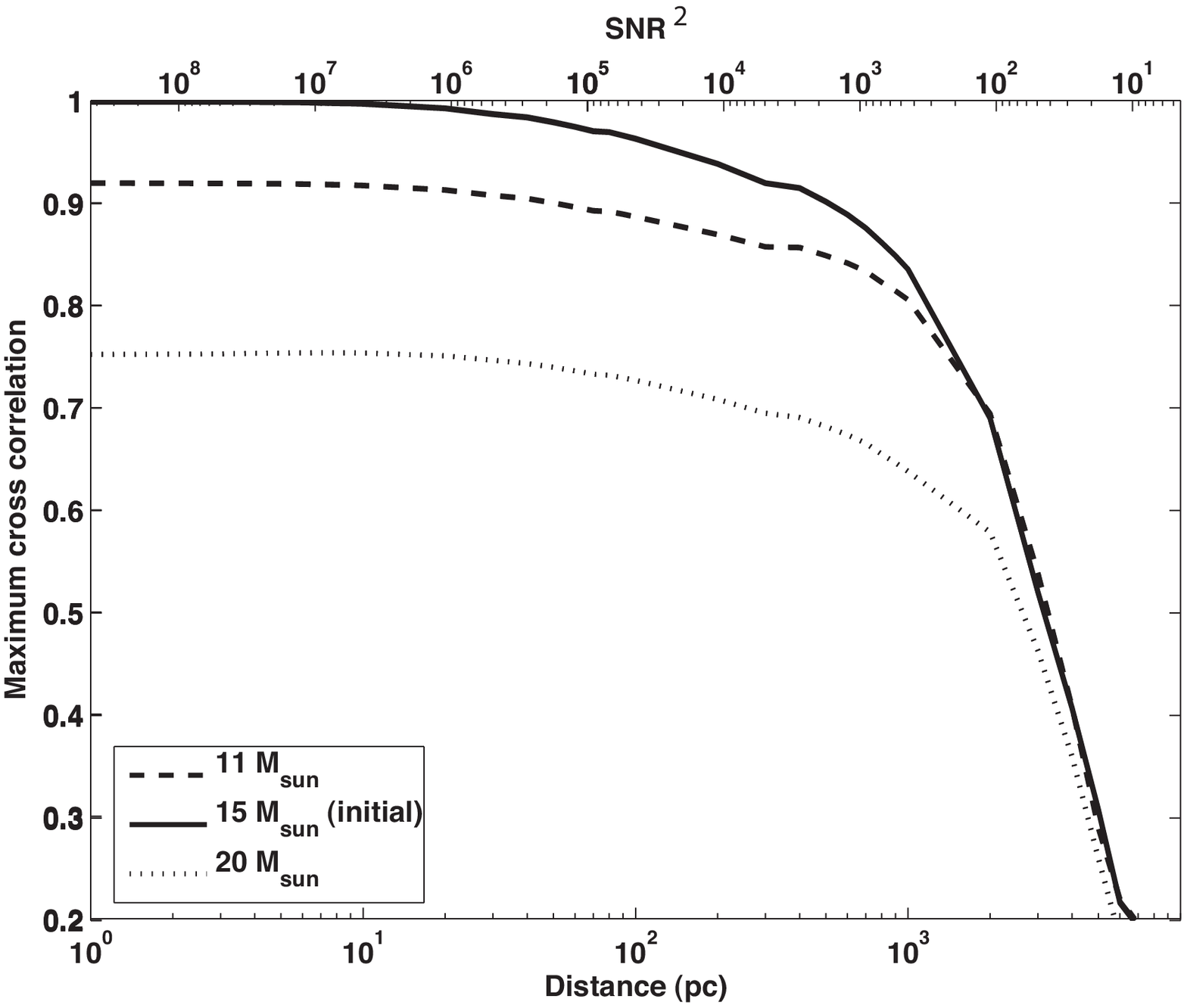}
\caption{Maximum cross-correlation between reconstructed waveforms
  and waveforms associated with models that differ only by progenitor
  mass versus core collapse distance and $\textrm{SNR}^2$.  The $\textrm{SNR}^2$ shown is
  the average for the two detectors.  The $\textrm{SNR}^2$ for the 4-km
  Hanford detector is 1.05 times that shown while the $\textrm{SNR}^2$ for the
  4-km Livingston detector is 0.95 times that shown.  The
  reconstructed waveform is inferred using maximum entropy from
  simulated detections that use a waveform from a model with a
  progenitor mass of 15 solar masses (Ott et al. [2004] model
  s15A1000B0.1) as the initial signal waveform as well as detector
  responses and noise levels from the fourth science run (S4). The
  inferred waveform is most similar to that generated by the model
  with the same progenitor mass for the simulations corresponding to
  core collapse events that occur less 
  than 2-3 kpc away.} \label{fig:mass}
\end{figure}

\begin{figure}
  \plotone{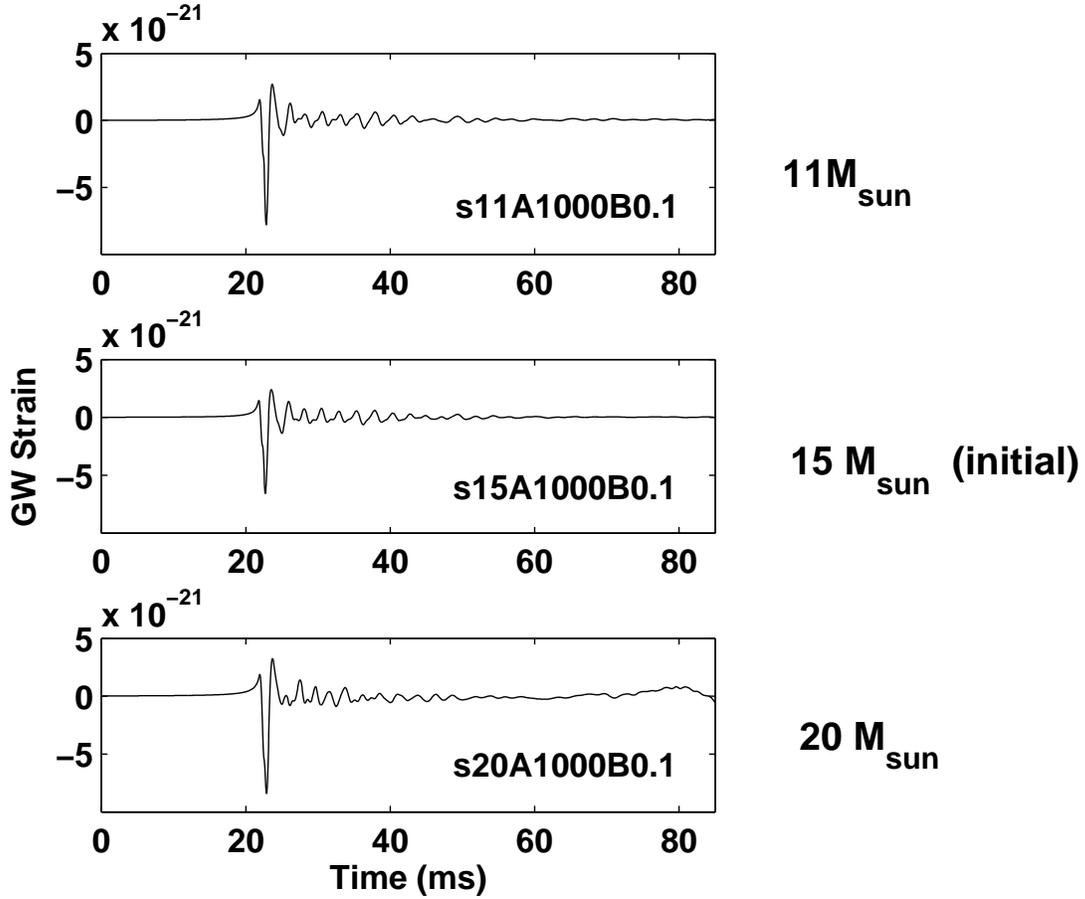}
\caption{Waveforms from models that differ only by progenitor
  mass.  The waveform corresponding to~\citet{ott:2004:gwf} model
  s15A1000B0.1 was used as the initial signal in the detection
  simulations.  The zero points of the time axes are chosen so that
  the minima of the waveforms occur at the same time for ease of
  comparison. The waveform amplitudes are scaled to correspond to
  core collapse events at 10 kpc.}\label{fig:wave_mass}
\end{figure}

Figure~\ref{fig:mass} shows the cross-correlation between the inferred
and actual waveforms associated with core collapse models that differ by
progenitor mass, but share the same rotational parameters. 
Figure~\ref{fig:wave_mass} shows the waveforms used
for the cross-correlations. Again, for S4 sensitivities the inferred
waveform most closely resembles the waveform from the model with the
same mass at distances up to 2-3~kpc.

\subsubsection{Rotation}\label{sec:Omega}
\begin{figure}
  \plotone{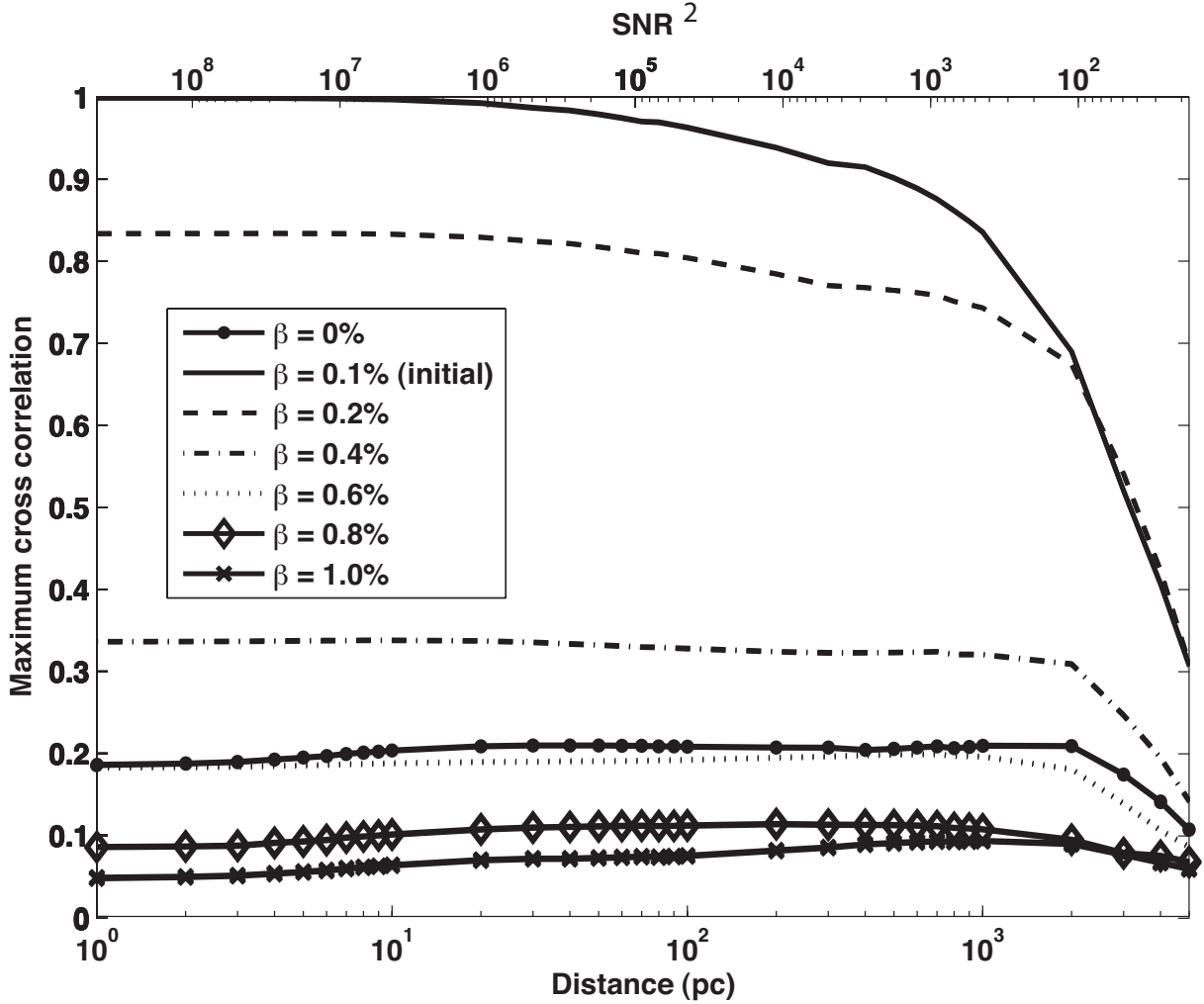}
\caption{Maximum cross-correlation between reconstructed waveforms
  and waveforms associated with models that differ only by rotation
  parameter $\beta$, which is defined in equation~(\ref{eq:beta}),
  versus core collapse distance and $\textrm{SNR}^2$.  The $\textrm{SNR}^2$ shown is the
  average for the two detectors.  The $\textrm{SNR}^2$ for the 4-km Hanford
  detector is 1.05 times that shown while the $\textrm{SNR}^2$ for the 4-km
  Livingston detector is 0.95 times that shown.  The reconstructed
  waveform is inferred using maximum entropy from simulated detections
  that use a waveform from a model with a rotation parameter of $\beta
  = 0.1\%$ (Ott et al.  [2004] model s15A1000B0.1) as the initial
  signal waveform as well as detector responses and noise levels from
  the fourth science run (S4).  The inferred waveform is most similar
  to that generated by the model with the same $\beta$ for the
  simulations corresponding to core collapse events
  that occur less than 2-3 kpc
  away.} \label{fig:beta}
\end{figure}

\begin{figure}
\plotone{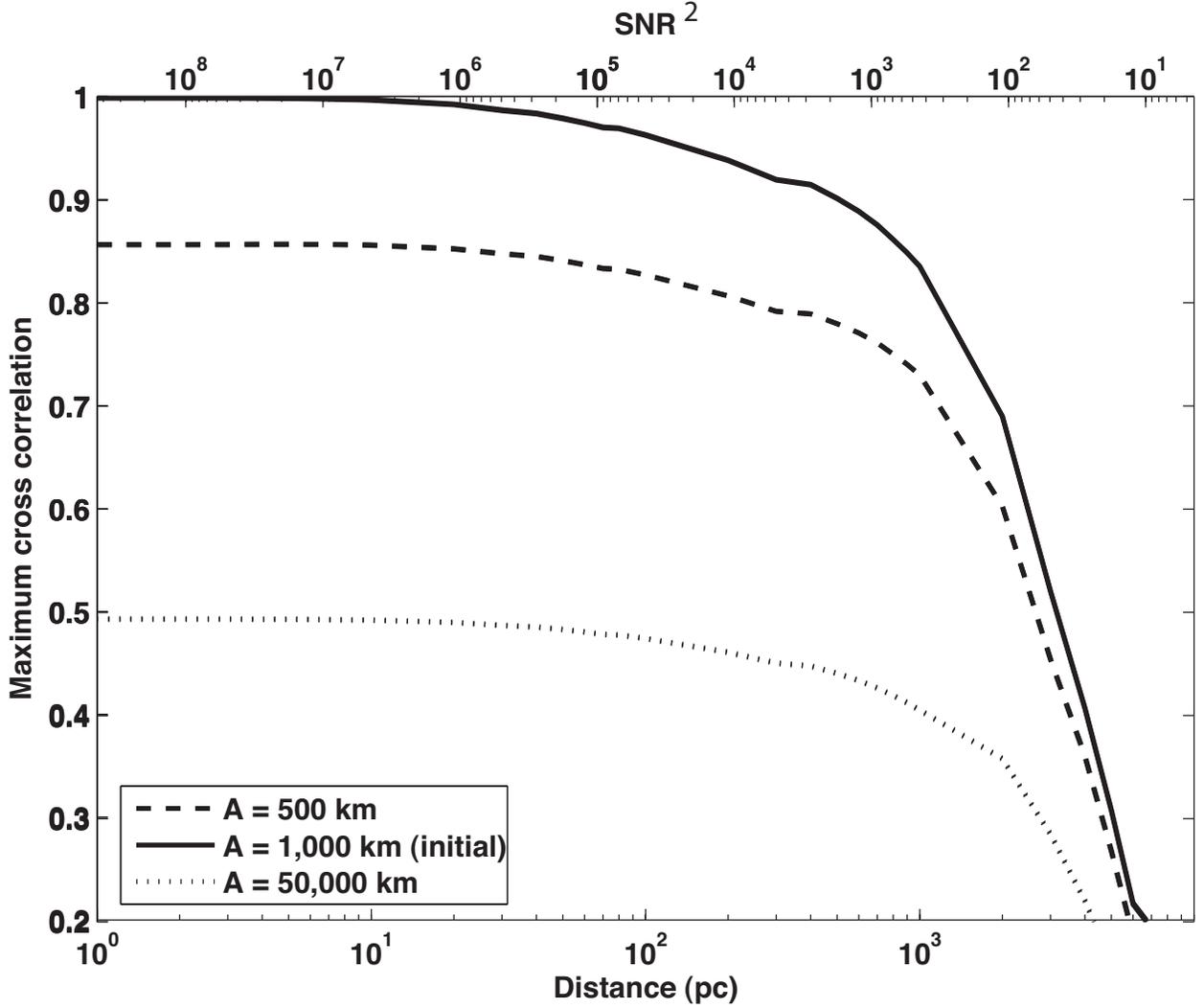}
\caption{Maximum cross-correlation between reconstructed waveforms
  and waveforms associated with models that differ only by initial
  degree of differential rotation as parameterized by $A$, which is
  defined in equation~(\ref{eq:A}), versus core collapse distance and
  $\textrm{SNR}^2$.  The $\textrm{SNR}^2$ shown is the average for the two detectors.
  The $\textrm{SNR}^2$ for the 4-km Hanford detector is 1.05 times that shown
  while the $\textrm{SNR}^2$ for the 4-km Livingston detector is 0.95 times
  that shown.  The reconstructed waveform is inferred using maximum
  entropy from simulated detections that use a waveform from a model
  with a differential rotation parameter of $A$ = 1,000 km (Ott et al.
  [2004] model s15A1000B0.1) as the initial signal waveform as well as
  detector responses and noise levels from the fourth science run
  (S4).  The inferred waveform is the most similar to that generated
  by the model with the same initial degree of differential rotation.}
\label{fig:diffrot}
\end{figure}

\begin{figure}
  \plotone{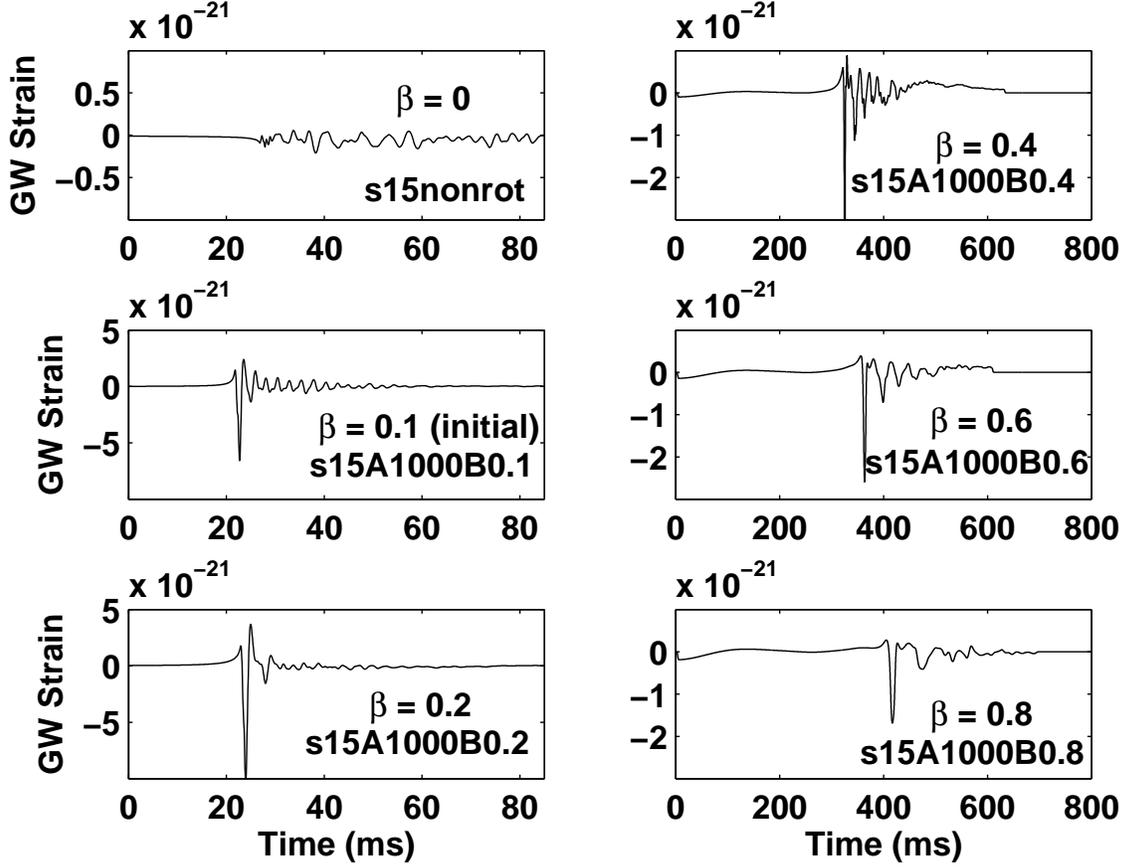}
\caption{Waveforms from models that differ only by rotation parameter 
  $\beta$, defined by equation~(\ref{eq:beta}).  The waveforms for
  larger $\beta$ ($\geq 0.4\%$) have significant amplitude over
  durations of hundreds of ms, while low $\beta$ waveforms last only
  for tens of ms.  The $\beta = 0\%$ waveform has very low amplitude
  as the non-rotating collapse is nearly spherically symmetric.  The
  waveform corresponding to~\cite{ott:2004:gwf} model s15A1000B0.1 was
  used as the initial signal in the detection simulations.  The zero
  point of the time axes for the plots on the right is chosen so that
  the onset of significant gravitational wave amplitude occurs at
  roughly the same time while the plots on the left show the first 800
  ms of the waveform.  The waveform amplitudes are scaled to
  correspond to core collapse events at 10 kpc.}\label{fig:wave_beta}
\end{figure}

\begin{figure}
\plotone{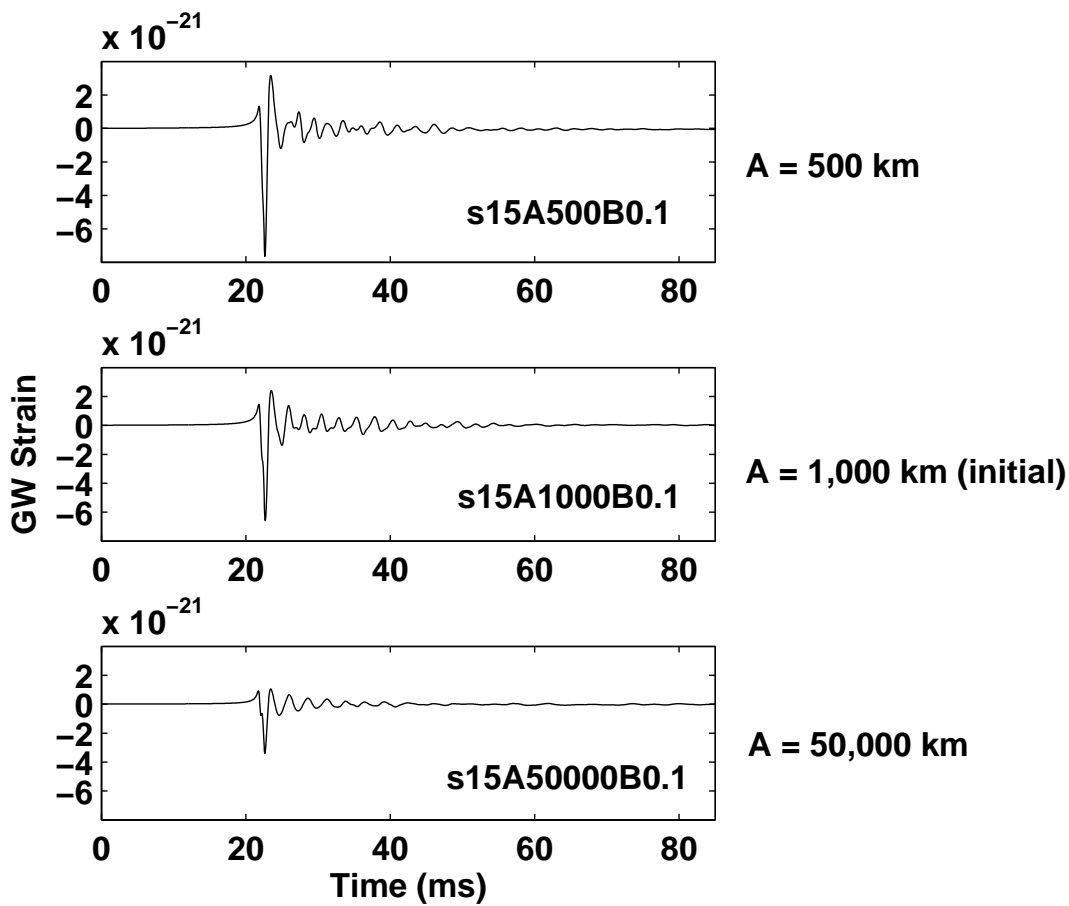}
\caption{Waveforms from models that differ only by initial degree of
  differential rotation.  The differential rotation parameter, $A$, is
  the distance at which the rotational velocity of the progenitor
  drops to half the rotational velocity at its center.  As $A$
  decreases the differential rotation of the progenitor becomes more
  extreme and the amplitudes of the gravitational waves increase.  The
  center plot shows the waveform corresponding to~\citet{ott:2004:gwf}
  model s15A1000B0.1 which was used as the initial signal in the
  detection simulations. The zero points of the time axes are chosen
  so that the minima of the waveforms occur at the same time for ease
  of comparison.  The waveform amplitudes are scaled to correspond to
  core collapse events at 10 kpc.}\label{fig:wave_A} 
\end{figure}

Figure~\ref{fig:beta} shows the cross-correlation between the inferred
waveform, the actual waveform, and the waveforms associated with a set
of models that differ only by rotation parameter $\beta$, while figure
\ref{fig:diffrot} shows the cross-correlations for models that differ
only by the differential rotation parameter $A$ (cf. eq. \ref{eq:A}).
Figures~\ref{fig:wave_beta} and~\ref{fig:wave_A} show the waveforms
for the set of models whose cross-correlations are shown in figures
\ref{fig:beta} and \ref{fig:diffrot}.  As expected, the
cross-correlation decreases as the rotational parameters of the models
depart from those associated with the actual waveform.

\section{DISCUSSION/CONCLUSIONS}
\label{sec:Conclude}

We have described and demonstrated the use of a Bayesian method, in the spirit of the so-called Maximum Entropy methods for analyzing astronomical image data, for inferring the time-dependent waveforms of the radiation incident on a network of gravitational wave detectors. In contrast to the methods developed for image analysis, which rely on the Principle of Maximum Entropy to select an a priori probability for use in the Bayesian analysis, we use the MaxEnt Principle to determine the appropriate choice of likelihood function for describing gravitational wave detector noise and use transformation group arguments to identify an appropriate, uninformative a priori probability density. We have demonstrated that the method described is quite capable of inferring the presence of signal in detector noise once the detector response to the signal begins to approach the noise rms, and very quickly converges on the correct signal amplitude and temporal structure of the incident signal as the actual signal amplitude increases. We have argued, based on the methods construction, a quantitative assessment of the correlation between the inferred and actual signal and the observed and expected signal-to-amplitudes associated of simulated data sets, and on a qualitative review of the residuals following subtraction of the detector networks response to the \emph{inferred} signal from the observations themselves, that the method developed here is the best that one can expect to be able to do for detector networks whose noise character is known only via its power spectral density and for signals that are assumed to have arbitrary time dependence. 

Inferring the gravitational wave signal incident on a set of detectors is but the first task of analysis. Our ultimate goal is to use the inferred signal to understand the radiation source. We demonstrate how this might be done using gravitational wave signatures calculated from the rotating core collapse supernovae models of \citet{ott:2004:gwf}. Using simulations drawn from \citet{ott:2004:gwf} we have generated simulated LIGO observations of core collapse supernovae, used our Bayesian analysis method to infer the most probable gravitational wave signal content of the data, and correlated this inferred signal with a wide range of signals drawn from the catalog of \citet{ott:2004:gwf}. We have found, in the context of this example, that at the
signal-to-noise ratios expected to be used as a threshold for detection of core-collapse supernova burst events in LIGO, the correlation is maximized when the inferred signal is matched against its parent from the \citet{ott:2004:gwf} catalog: thus, the analysis method we describe here infers the incident gravitational wave signal with enough fidelity to distinguish between models of different maximum central density, angular momentum, differential rotation, and progenitor mass.

To make the comparison between the inferred and model waveforms we
used the maximum of the cross-correlation between the two,
choosing the model whose maximum cross-correlation with the inferred
waveform is the greatest. A Bayesian treatment that assigns odds to
different models given the inferred waveform is a natural next step;
however, we note that a Bayesian treatment would involve the
exponential of the cross-correlation function (as opposed to the
maximum cross-correlation) marginalized over the lag. When the
cross-correlation is a sharply peaked function, this will be
proportional to the exponential of the maximum of the
cross-correlation among the different models. Thus, we conclude that
the cross-correlation as we use it here is almost certainly likely to
choose the same model as a more sophisticated Bayesian analysis,
except in those cases where the waveforms of the different models are
nearly indistinguishable.

While our example applications have been in terms of two
interferometers, there is nothing in the description of the method
that specifies the number of detectors and the application to three or
more detectors is no different than for one or two detectors. The 
example application provided here involved just the two main LIGO detectors; 
however, we could just as well have augmented the analysis by including
simulated data Virgo, GEO, TAMA and the operating bar detectors. 
For the particular simulations described here, adding any one of these
detectors would most likely not improve our estimates of
the signal since --- for the examples presented here --- we chose to place
the source in the most favorable sky location and orientation for detection 
by the LIGO Hanford Observatory (LHO) site.  For a source so
positioned the strain amplitude in the LIGO Livingston detector is
89\% of that for the LHO detectors, while for the GEO600 site detector
it is 42\%, 22\% for TAMA300, and 1\% for Virgo. On the other hand,
the addition of these other detectors to a real network would increase
the overall sensitivity across the sky, as a source optimally oriented
for Virgo would, by reciprocity, be very poorly oriented for the LHO
detectors.

Gravitational wave astronomy is not yet a reality. Nevertheless, with
the detectors now operating and the enhancements currently underway,
it is only a matter of time before we have observations of from these
instruments, from which we can gain insight into astronomical
phenomena that are otherwise hidden from our sight. Methods like those
described here will allow us to move beyond gravitational wave
detection and realize the full promise of gravitational wave astronomy.

\acknowledgments 

We are grateful to our anonymous referee for helpful recommendations, 
which have helped us significantly strengthen this paper's accessibility and 
relevance. 
TZS and LSF gratefully acknowledge informative discussions with Graham
Woan, Malik Rakhmanov, Nelson Christensen, and Soumya Mohanty.
TZS and LSF were supported by NSF through grant PHY-00-99559 and the
Center for Gravitational Wave Physics. LSF also acknowledges the
support of NSF through awards PHY 06-53462 and PHY 05-55615, NASA through award
NNG05GF71G, and the Penn State Center for Space Research Programs.
AB acknowledges support for this work from the Scientific Discovery
through Advanced Computing (SciDAC) program of the DOE, grant numbers
DE-FC02-01ER41184 and DE-FC02-06ER41452 and from the NSF under grant number AST-0504947.
CDO acknowleges support through a Joint Institute for Nuclear
  Astrophysics postdoctoral fellowship, sub-award no.~61-5292UA of NFS
  award no.~86-6004791. The core-collapse calculations were carried
  out on the local Arizona cluster, at the Albert Einstein Institute,
  and at the National Center for Supercomputing Applications (NCSA) 
  under Teragrid computer time grant TG-MCA02N014. The Center for Gravitational Wave Physics is funded by the National
Science Foundation via cooperative agreement PHY 01-14375.


\begin{thebibliography}{74}
\expandafter\ifx\csname natexlab\endcsname\relax\def\natexlab#1{#1}\fi

\bibitem[{Abbott {et~al.}(2004{\natexlab{a}})Abbott, Abbott, Adhikari, Ageev,
  Allen, Amin, Anderson, Anderson, Araya, Armandula, Asiri, Aufmuth, Aulbert,
  Babak, Balasubramanian, Ballmer, Barish, Barker, {Barker-Patton}, Barnes,
  Barr, Barton, Bayer, Beausoleil, Belczynski, Bennett, Berukoff, Betzwieser,
  Bhawal, Bilenko, Billingsley, Black, Blackburn, {Bland-Weaver}, Bochner,
  Bogue, Bork, Bose, Brady, Braginsky, Brau, Brown, Brozek, Bullington,
  Buonanno, Burgess, Busby, Butler, Byer, Cadonati, Cagnoli, Camp, Cantley,
  Cardenas, Carter, Casey, Castiglione, Chandler, Chapsky, Charlton, Chatterji,
  Chen, Chickarmane, Chin, Christensen, Churches, Colacino, Coldwell, Coles,
  Cook, Corbitt, Coyne, Creighton, Creighton, Crooks, Csatorday, Cusack,
  Cutler, {D'Ambrosio}, Danzmann, Davies, Daw, Debra, Delker, Desalvo,
  Dhurandar, {D{\'\i}az}, Ding, Drever, Dupuis, Ebeling, Edlund, Ehrens,
  Elliffe, Etzel, Evans, Evans, Fallnich, Farnham, Fejer, Fine, Finn, Flanagan,
  Freise, Frey, Fritschel, Frolov, Fyffe, Ganezer, Giaime, Gillespie, Goda,
  {Gonz{\'a}lez}, {Go \ss ler}, {Grandcl{\'e}ment}, Grant, Gray, Gretarsson,
  Grimmett, Grote, Grunewald, Guenther, Gustafson, Gustafson, Hamilton,
  Hammond, Hanson, Hardham, Harry, Hartunian, Heefner, Hefetz, Heinzel, Heng,
  Hennessy, Hepler, Heptonstall, Heurs, Hewitson, Hindman, Hoang, Hough,
  Hrynevych, Hua, Ingley, Ito, Itoh, Ivanov, Jennrich, Johnson, Johnston,
  Jones, Jungwirth, Kalogera, Katsavounidis, Kawabe, Kawamura, Kells, Kern,
  Khan, Killbourn, Killow, Kim, King, King, Klimenko, Kloevekorn, Koranda,
  {K{\"o}tter}, Kovalik, Kozak, Krishnan, Landry, Langdale, Lantz, Lawrence,
  Lazzarini, Lei, Leonhardt, Leonor, Libbrecht, Lindquist, Liu, Logan, Lormand,
  Lubinski, {L{\"u}ck}, Lyons, Machenschalk, Macinnis, Mageswaran, Mailand,
  Majid, Malec, Mann, Marin, {M{\'a}rka}, Maros, Mason, Mason, Matherny,
  Matone, Mavalvala, McCarthy, McClelland, McHugh, McNamara, Mendell, Meshkov,
  Messenger, Mitrofanov, Mitselmakher, Mittleman, Miyakawa, Miyoki, Mohanty,
  Moreno, Mossavi, Mours, Mueller, Mukherjee, Myers, Nagano, Nash, Naundorf,
  Nayak, Newton, Nocera, Nutzman, Olson, {O'Reilly}, {J.~Ottaway}, Ottewill,
  Ouimette, Overmier, Owen, Papa, Parameswariah, Parameswariah, Pedraza, Penn,
  Pitkin, Plissi, Pratt, Quetschke, Raab, Radkins, Rahkola, Rakhmanov, Rao,
  Redding, Regehr, Regimbau, Reilly, Reithmaier, Reitze, Richman, Riesen,
  Riles, Rizzi, Robertson, Robertson, Robison, Roddy, Rollins, Romano, Romie,
  Rong, Rose, Rotthoff, Rowan, {R{\"u}diger}, Russell, Ryan, Salzman, Sanders,
  Sannibale, Sathyaprakash, Saulson, Savage, Sazonov, Schilling, Schlaufman,
  Schmidt, Schofield, Schrempel, Schutz, Schwinberg, Scott, Searle, Sears,
  Seel, Sengupta, Shapiro, Shawhan, Shoemaker, Shu, Sibley, Siemens, Sievers,
  Sigg, Sintes, Skeldon, Smith, Smith, Smith, Sneddon, Spero, Stapfer, Strain,
  Strom, Stuver, Summerscales, Sumner, Sutton, Sylvestre, Takamori, Tanner,
  Tariq, Taylor, Taylor, Thorne, Tibbits, Tilav, Tinto, Tokmakov, Torres,
  Torrie, Traeger, Traylor, Tyler, Ugolini, Vallisneri, {van Putten}, Vass,
  Vecchio, Vorvick, Vyatchanin, Wallace, Walther, Ward, Ware, Watts, Webber,
  Weidner, Weiland, Weinstein, Weiss, Welling, Wen, Wen, Whelan, Whitcomb,
  Whiting, Willems, Williams, Williams, Willke, Wilson, Winjum, Winkler, Wise,
  Wiseman, Woan, Wooley, Worden, Yakushin, Yamamoto, Yoshida, Zawischa, Zhang,
  Zotov, Zucker, \& Zweizig}]{abbott:2004:dda}
Abbott, B., Abbott, R., Adhikari, R., Ageev, A., Allen, B., Amin, R., Anderson,
  S.~B., Anderson, W.~G. et al.
  2004{\natexlab{a}}, Nucl. Instrum. Methods Phys. Research, 517, 154

\bibitem[{Abbott {et~al.}(2004{\natexlab{b}})Abbott, Abbott, Adhikari, Ageev,
  Allen, Amin, Anderson, Anderson, Araya, Armandula, Asiri, Aufmuth, Aulbert,
  Babak, Balasubramanian, Ballmer, Barish, Barker, {Barker-Patton}, Barnes,
  Barr, Barton, Bayer, Beausoleil, Belczynski, Bennett, Berukoff, Betzwieser,
  Bhawal, Bilenko, Billingsley, Black, Blackburn, {Bland-Weaver}, Bochner,
  Bogue, Bork, Bose, Brady, Braginsky, Brau, Brown, Brozek, Bullington,
  Buonanno, Burgess, Busby, Butler, Byer, Cadonati, Cagnoli, Camp, Cantley,
  Cardenas, Carter, Casey, Castiglione, Chandler, Chapsky, Charlton, Chatterji,
  Chen, Chickarmane, Chin, Christensen, Churches, Colacino, Coldwell, Coles,
  Cook, Corbitt, Coyne, Creighton, Creighton, Crooks, Csatorday, Cusack,
  Cutler, {D'Ambrosio}, Danzmann, Davies, Daw, Debra, Delker, Desalvo,
  Dhurandhar, {D{\'\i}az}, Ding, Drever, Dupuis, Ebeling, Edlund, Ehrens,
  Elliffe, Etzel, Evans, Evans, Fallnich, Farnham, Fejer, Fine, Finn, Flanagan,
  Freise, Frey, Fritschel, Frolov, Fyffe, Ganezer, Giaime, Gillespie, Goda,
  {Gonz{\'a}lez}, {Go \ss ler}, {Grandcl{\'e}ment}, Grant, Gray, Gretarsson,
  Grimmett, Grote, Grunewald, Guenther, Gustafson, Gustafson, Hamilton,
  Hammond, Hanson, Hardham, Harry, Hartunian, Heefner, Hefetz, Heinzel, Heng,
  Hennessy, Hepler, Heptonstall, Heurs, Hewitson, Hindman, Hoang, Hough,
  Hrynevych, Hua, Ingley, Ito, Itoh, Ivanov, Jennrich, Johnson, Johnston,
  Jones, Jungwirth, Kalogera, Katsavounidis, Kawabe, Kawamura, Kells, Kern,
  Khan, Killbourn, Killow, Kim, King, King, Klimenko, Kloevekorn, Koranda,
  {K{\"o}tter}, Kovalik, Kozak, Krishnan, Landry, Langdale, Lantz, Lawrence,
  Lazzarini, Lei, Leonhardt, Leonor, Libbrecht, Lindquist, Liu, Logan, Lormand,
  Lubinski, {L{\"u}ck}, Lyons, Machenschalk, Macinnis, Mageswaran, Mailand,
  Majid, Malec, Mann, Marin, {M{\'a}rka}, Maros, Mason, Mason, Matherny,
  Matone, Mavalvala, McCarthy, McClelland, McHugh, McNamara, Mendell, Meshkov,
  Messenger, Mitrofanov, Mitselmakher, Mittleman, Miyakawa, Miyoki, Mohanty,
  Moreno, Mossavi, Mours, Mueller, Mukherjee, Myers, Nagano, Nash, Naundorf,
  Nayak, Newton, Nocera, Nutzman, Olson, {O'Reilly}, Ottaway, Ottewill,
  Ouimette, Overmier, Owen, Papa, Parameswariah, Parameswariah, Pedraza, Penn,
  Pitkin, Plissi, Pratt, Quetschke, Raab, Radkins, Rahkola, Rakhmanov, Rao,
  Redding, Regehr, Regimbau, Reilly, Reithmaier, Reitze, Richman, Riesen,
  Riles, Rizzi, Robertson, Robertson, Robison, Roddy, Rollins, Romano, Romie,
  Rong, Rose, Rotthoff, Rowan, {R{\"u}diger}, Russell, Ryan, Salzman, Sanders,
  Sannibale, Sathyaprakash, Saulson, Savage, Sazonov, Schilling, Schlaufman,
  Schmidt, Schofield, Schrempel, Schutz, Schwinberg, Scott, Searle, Sears,
  Seel, Sengupta, Shapiro, Shawhan, Shoemaker, Shu, Sibley, Siemens, Sievers,
  Sigg, Sintes, Skeldon, Smith, Smith, Smith, Sneddon, Spero, Stapfer, Strain,
  Strom, Stuver, Summerscales, Sumner, Sutton, Sylvestre, Takamori, Tanner,
  Tariq, Taylor, Taylor, Thorne, Tibbits, Tilav, Tinto, Tokmakov, Torres,
  Torrie, Traeger, Traylor, Tyler, Ugolini, Vallisneri, {van Putten}, Vass,
  Vecchio, Vorvick, Vyachanin, Wallace, Walther, Ward, Ware, Watts, Webber,
  Weidner, Weiland, Weinstein, Weiss, Welling, Wen, Wen, Whelan, Whitcomb,
  Whiting, Willems, Williams, Williams, Willke, Wilson, Winjum, Winkler, Wise,
  Wiseman, Woan, Wooley, Worden, Yakushin, Yamamoto, Yoshida, Zawischa, Zhang,
  Zotov, Zucker, \& Zweizig}]{abbott:2004:aol}
Abbott, B., Abbott, R., Adhikari, R., Ageev, A., Allen, B., Amin, R., Anderson,
  S.~B., Anderson, W.~G. et al.
  2004{\natexlab{b}}, Phys. Rev. D, 69, 122001

\bibitem[{Acernese {et~al.}(2006)Acernese, Amico, Alshourbagy, Antonucci,
  Aoudia, Avino, Babusci, Ballardin, Barone, Barsotti, Barsuglia, Beauville,
  Bigotta, Birindelli, Bizouard, Boccara, Bondu, Bosi, Bradaschia, Braccini,
  Brillet, Brisson, Brocco, Buskulic, Calloni, Campagna, Cavalier, Cavalieri,
  Cella, Cesarini, Chassande-Mottin, Corda, Cottone, Clapson, Cleva, Coulon,
  Cuoco, Dari, Dattilo, Davier, Rosa, Fiore, Virgilio, Dujardin, Eleuteri,
  Enard, Ferrante, Fidecaro, Fiori, Flaminio, Fournier, Francois, Frasca,
  Frasconi, Freise, Gammaitoni, Garufi, Gennai, Giazotto, Giordano, Giordano,
  Gouaty, Grosjean, Guidi, Hebri, Heitmann, Hello, Holloway, Karkar,
  Kreckelbergh, Penna, Laval, Leroy, Letendre, Lorenzini, Loriette, Loupias,
  Losurdo, Mackowski, Majorana, Man, Mantovani, Marchesoni, Marion, Marque,
  Martelli, Masserot, Mazzoni, Milano, Moins, Moreau, Morgado, Mours, Pai,
  Palomba, Paoletti, Pardi, Pasqualetti, Passaquieti, Passuello, Perniola,
  Piergiovanni, Pinard, Poggiani, Punturo, Puppo, Qipiani, Rapagnani, Reita,
  Remillieux, Ricci, Ricciardi, Ruggi, Russo, Solimeno, Spallicci, Stanga,
  Taddei, Tonelli, Toncelli, Tournefier, Travasso, Vajente, Verkindt, Vetrano,
  Vicer\'{e}, Vinet, Vocca, Yvert, \& Zhang}]{acernese:2006:vs}
Acernese, F., Amico, P., Alshourbagy, M., Antonucci, F., Aoudia, S., Avino, S.,
  Babusci, D., Ballardin, G. et al. 2006, Class. Quantum Grav., 23, S635

\bibitem[{Adhikari {et~al.}(2003)Adhikari, Gonz{\'a}lez, Landry, \&
  O'Reilly}]{adhikari:2003:col}
Adhikari, R., Gonz{\'a}lez, G., Landry, M., \& O'Reilly, B. 2003, Class.
  Quantum Grav., 20, S903

\bibitem[{Barreiro {et~al.}(2004)Barreiro, Hobson, Banday, Lasenby, Stolyarov,
  Vielva, \& {G{\'o}rski}}]{barreiro:2004:fsu}
Barreiro, R.~B., Hobson, M.~P., Banday, A.~J., Lasenby, A.~N., Stolyarov, V.,
  Vielva, P., \& {G{\'o}rski}, K.~M. 2004, {MNRAS}, 351, 515

\bibitem[{Barreiro {et~al.}(2001)Barreiro, Vielva, Hobson,
  {Mart{\'\i}nez-Gonz{\'a}lez}, Lasenby, Sanz, \&
  Toffolatti}]{barreiro:2001:rms}
Barreiro, R.~B., Vielva, P., Hobson, M.~P., {Mart{\'\i}nez-Gonz{\'a}lez}, E.,
  Lasenby, A.~N., Sanz, J.~L., \& Toffolatti, L. 2001, in Mining the Sky, ed. A.~J.~Bandy, S.~Zaroubi \& M.~Bartelmann (Berlin: Springer-Verlag), 465

\bibitem[{Bennett {et~al.}(2003)Bennett, Halpern, Hinshaw, Jarosik, Kogut,
  Limon, Meyer, Page, Spergel, Tucker, Wollack, Wright, Barnes, Greason, Hill,
  Komatsu, Nolta, Odegard, Peiris, Verde, \& Weiland}]{bennett:2003:fwm}
Bennett, C.~L., Halpern, M., Hinshaw, G., Jarosik, N., Kogut, A., Limon, M.,
  Meyer, S.~S., Page, L. et al. 2003, ApJS, 148, 1

\bibitem[{Bretthorst(1988)}]{bretthorst:1988:bsa}
Bretthorst, G.~L. 1988, in Lecture Notes in Statistics, Vol.~48, Bayesian
  Spectrum Analysis and Parameter Estimation, ed. J.~Berger, S.~Fienberg,
  J.~Gani, K.~Krickeberg, \& B.~Singer (New York: Springer-Verlag)

\bibitem[{Burrows \& Hayes(1996)}]{burrows:1996:pra}
Burrows, A., \& Hayes, J. 1996, Phys. Rev. Lett., 76, 352

\bibitem[{Burrows {et~al.}(2006)Burrows, Livne, Dessart, Ott, \&
  Murphy}]{burrows:2006:nmf}
Burrows, A., Livne, E., Dessart, L., Ott, C.~D., \& Murphy, J. 2006, ApJ, 640,
  878

\bibitem[{Burrows {et~al.}(2007)Burrows, Livne, Dessart, Ott, \&
  Murphy}]{burrows:2007:foa}
---. 2007, ApJ, 655, 416

\bibitem[{Cox(1946)}]{cox:1946:pfa}
Cox, R.~T. 1946, Am. Jour. Phys., 14, 1

\bibitem[{Cox(1961)}]{cox:1961:aop}
---. 1961, The Algebra of Probable Inference (Baltimore: Johns Hopkins Press)

\bibitem[{Dimmelmeier {et~al.}(2002)Dimmelmeier, Font, \&
  {M{\"u}ller}}]{dimmelmeier:2002:rso}
Dimmelmeier, H., Font, J.~A., \& {M{\"u}ller}, E. 2002, A\&A, 393, 523

\bibitem[{Dimmelmeier {et~al.}(2007{\natexlab{a}})Dimmelmeier, Ott, Janka,
  Marek, \& Mueller}]{dimmelmeier:2007:ggs}
Dimmelmeier, H., Ott, C.~D., Janka, H.-T., Marek, A., \& Mueller, E.
  2007{\natexlab{a}}, ArXiv e-prints, 705

\bibitem[{Dimmelmeier {et~al.}(2007{\natexlab{b}})Dimmelmeier, Ott, Janka,
  Marek, \& {M{\"u}ller}}]{dimmelmeier:2007:ggs:1}
Dimmelmeier, H., Ott, C.~D., Janka, H.-T., Marek, A., \& {M{\"u}ller}, E.
  2007{\natexlab{b}}, Phys. Rev. Lett., 98, 251101

\bibitem[{Evans \& Stark(2002)}]{evans:2002:ipa}
Evans, S.~N., \& Stark, P.~B. 2002, Inverse Problems, 18, R55

\bibitem[{Ferrari {et~al.}(2003)Ferrari, Miniutti, \& Pons}]{ferrari:2003:gwf}
Ferrari, V., Miniutti, G., \& Pons, J.~A. 2003, Class. Quantum Grav., 20, 841

\bibitem[{Finn(1991)}]{finn:1991:dog}
Finn, L.~S. 1991, Annals Of The New York Academy Of Sciences, 631, 156

\bibitem[{Finn(1992)}]{finn:1992:dma}
---. 1992, Phys. Rev. D, 46, 5236

\bibitem[{Finn \& Chernoff(1993)}]{finn:1993:obi}
Finn, L.~S., \& Chernoff, D.~F. 1993, Phys. Rev. D, 47, 2198

\bibitem[{Fryer {et~al.}(2002)Fryer, Holz, \& Hughes}]{fryer:2002:gwe}
Fryer, C.~L., Holz, D.~E., \& Hughes, S.~A. 2002, ApJ, 565, 430

\bibitem[{{Gonz\'alez} {et~al.}(2004){Gonz\'alez}, Landry, {O'Reilly}, \&
  Radkins}]{gonzalez:2004:col}
{Gonz\'alez}, G., Landry, M., {O'Reilly}, B., \& Radkins, H. 2004, Calibration
  of the {LIGO} detectors for {S2}, Tech. Rep. T040060--01-D, Laser
  Interferometer Gravitational Wave Observatory, {LIGO} Document Control
  Center, California Institute of Technology, available at
  \texttt{http://admdbsrv.ligo.caltech.edu/dcc}

\bibitem[{Gull \& Daniell(1978)}]{gull:1978:irf}
Gull, S.~F., \& Daniell, G.~J. 1978, Nature, 272, 686

\bibitem[{Gull \& Daniell(1979)}]{gull:1979:mem}
Gull, S.~F. \& Daniell, G.~J. 1979, in Astrophysics and Space Science Library,
  Vol.~76, Image Formation from Coherence Functions in Astronomy, ed. C.~van
  Schooneveld, IAU (C. D. Reidel), 219

\bibitem[{G{\"u}rsel \& Tinto(1989)}]{gursel:1989:nos}
G{\"u}rsel, Y., \& Tinto, M. 1989, Phys. Rev. D, 40, 3884

\bibitem[{Heger {et~al.}(2000)Heger, Langer, \& Woosley}]{heger:2000:peo}
Heger, A., Langer, N., \& Woosley, S.~E. 2000, ApJ, 528, 368

\bibitem[{{Heger} {et~al.}(2004){Heger}, {Woosley}, {Langer}, \&
  {Spruit}}]{heger:2004:peo}
{Heger}, A., {Woosley}, S.~E., {Langer}, N., \& {Spruit}, H.~C. 2004, in IAU
  Symposium, Vol. 215, Stellar Rotation, ed. A.~{Maeder} \& P.~{Eenens} (San
  Francisco: Astronomical Society of the Pacific), 591--+

\bibitem[{Jaynes(1957{\natexlab{a}})}]{jaynes:1957:ita:1}
Jaynes, E.~T. 1957{\natexlab{a}}, Phys. Rev., 106, 620

\bibitem[{Jaynes(1957{\natexlab{b}})}]{jaynes:1957:ita}
---. 1957{\natexlab{b}}, Phys. Rev., 108, 171

\bibitem[{Jaynes(1968)}]{jaynes:1968:pp}
---. 1968, IEEE Transactions on Systems Science and Cybernetics, 4, 227

\bibitem[{Jones {et~al.}(1998)Jones, Hancock, Lasenby, Davies, Gutierrez,
  Rocha, Watson, \& Rebolo}]{jones:1998:1tc}
Jones, A.~W., Hancock, S., Lasenby, A.~N., Davies, R.~D., Gutierrez, C.~M.,
  Rocha, G., Watson, R.~A., \& Rebolo, R. 1998, {MNRAS}, 294, 582

\bibitem[{Jones {et~al.}(1999)Jones, Lasenby, Mukherjee, Gutierrez, Davies,
  Watson, Hoyland, \& Rebolo}]{jones:1999:mmj}
Jones, A.~W., Lasenby, A.~N., Mukherjee, P., Gutierrez, C.~M., Davies, R.~D.,
  Watson, R.~A., Hoyland, R., \& Rebolo, R. 1999, {MNRAS}, 310, 105

\bibitem[{Kittel(1958)}]{kittel:1958:esp}
Kittel, C. 1958, Elementary Statistical Physics (New York: Wiley)

\bibitem[{Lattimer \& Swesty(1991)}]{lattimer:1991:geo}
Lattimer, J.~M., \& Swesty, F.~D. 1991, Nucl.~Phys~A, 535, 331

\bibitem[{Lazzarini(2002)}]{lazzarini:2002:lss}
Lazzarini, A. 2002, {LIGO} S1 Strain Sensitivity, Tech. Rep. G020461--01-E,
  Laser Interferometer Gravitational Wave Observatory, {LIGO} Document Control
  Center, California Institute of Technology, available at
  \texttt{http://admdbsrv.ligo.caltech.edu/dcc}

\bibitem[{Lazzarini(2003)}]{lazzarini:2003:lss}
---. 2003, {LIGO} S2 Strain Sensitivity, Tech. Rep. G030379--00-E, Laser
  Interferometer Gravitational Wave Observatory, {LIGO} Document Control
  Center, California Institute of Technology, available at
  \texttt{http://admdbsrv.ligo.caltech.edu/dcc}

\bibitem[{Lazzarini(2004)}]{lazzarini:2004:sbs}
---. 2004, {S3} Best Strain Sensitivities, Tech. Rep. G040023--00-E, Laser
  Interferometer Gravitational Wave Observatory, {LIGO} Document Control
  Center, California Institute of Technology, available at
  \texttt{http://admdbsrv.ligo.caltech.edu/dcc}

\bibitem[{Lazzarini(2005)}]{lazzarini:2005:sss}
---. 2005, {S4} Strain Sensitivities for the {LIGO} Inteferometers, Tech. Rep.
  G050230--02-E, Laser Interferometer Gravitational Wave Observatory, {LIGO}
  Document Control Center, California Institute of Technology, available at
  \texttt{http://admdbsrv.ligo.caltech.edu/dcc}

\bibitem[{LIGO(2007)}]{ligo:2007:web}
LIGO Laboratory. 2007, \texttt{http://ligo.caltech.edu}
  
\bibitem[{{L{\"u}ck} {et~al.}(2006){L{\"u}ck}, Hewitson, Ajith, Allen, Aufmuth,
  Aulbert, Babak, Balasubramanian, Barr, Berukoff, Bunkowski, Cagnoli, Cantley,
  Casey, Chelkowski, Chen, Churches, Cokelaer, Colacino, Crooks, Cutler,
  Danzmann, Dupuis, Elliffe, Fallnich, Franzen, Freise, Gholami, {Go \ss ler},
  Grant, Grote, Grunewald, Harms, Hage, Heinzel, Heng, Hepstonstall, Heurs,
  Hild, Hough, Itoh, Jones, Jones, Huttner, {K{\"o}tter}, Krishnan, Kwee, Luna,
  Machenschalk, Malec, Mercer, Meier, Messenger, Mohanty, Mossavi, Mukherjee,
  Murray, Newton, Papa, {Perreur-Lloyd}, Pitkin, Plissi, Prix, Quetschke, Re,
  Regimbau, Rehbein, Reid, Ribichini, Robertson, Robertson, Robinson, Romano,
  Rowan, {R{\"u}diger}, Sathyaprakash, Schilling, Schnabel, Schutz, Seifert,
  Sintes, Smith, Sneddon, Strain, Taylor, Taylor, {Th{\"u}ring}, Ungarelli,
  Vahlbruch, Vecchio, Veitch, Ward, Weiland, Welling, Wen, Williams, Willke,
  Winkler, Woan, \& Zhu}]{luck:2006:sog}
{L{\"u}ck}, H., Hewitson, M., Ajith, P., Allen, B., Aufmuth, P., Aulbert, C.,
  Babak, S., Balasubramanian, R. et al. 2006, Class. Quantum Grav., 23, 71

\bibitem[{Maisinger {et~al.}(1997)Maisinger, Hobson, \&
  Lasenby}]{maisinger:1997:mem}
Maisinger, K., Hobson, M.~P., \& Lasenby, A.~N. 1997, {MNRAS}, 290, 313

\bibitem[{Maisinger {et~al.}(2004)Maisinger, Hobson, \&
  Lasenby}]{maisinger:2004:mir}
---. 2004, {MNRAS}, 347, 339

\bibitem[{Mandic(2006)}]{mandic:2006:sol}
Mandic, V. 2006, PoS, HEP2005, 037

\bibitem[{Misner {et~al.}(1973)Misner, Thorne, \& Wheeler}]{misner:1973:g}
Misner, C.~W., Thorne, K.~S., \& Wheeler, J.~A. 1973, Gravitation (San
  Francisco: W. H. Freeman \& Co.)

\bibitem[{{M{\"u}ller} {et~al.}(2004){M{\"u}ller}, Rampp, Buras, Janka, \&
  Shoemaker}]{muller:2004:tgw}
{M{\"u}ller}, E., Rampp, M., Buras, R., Janka, H.-T., \& Shoemaker, D.~H. 2004,
  ApJ, 603, 221

\bibitem[{Narayan \& Nityananda(1984)}]{narayan:1984:me-}
Narayan, R., \& Nityananda, R. 1984, in Indirect Imaging. Measurement and
  Processing for Indirect Imaging, ed. J.~A. Roberts (Cambridge, England: Cambridge University Press), 281

\bibitem[{Narayan \& Nityananda(1986)}]{narayan:1986:mei}
Narayan, R., \& Nityananda, R. 1986, ARA\&A, 24, 127

\bibitem[{Neumaier(1998)}]{neumaier:1998:sia}
Neumaier, A. 1998, SIAM Review, 40, 636

\bibitem[{Nityananda \& Narayan(1982)}]{nityananda:1982:mei}
Nityananda, R., \& Narayan, R. 1982, Journal of Astrophysics and Astronomy, 3,
  419

\bibitem[{Nityananda \& Narayan(1983)}]{nityananda:1983:rop}
---. 1983, A\&A, 118, 194

\bibitem[{Ott {et~al.}(2006)Ott, Burrows, Dessart, \& Livne}]{ott:2006:nmf}
Ott, C.~D., Burrows, A., Dessart, L., \& Livne, E. 2006, Phys. Rev. Lett., 96,
  201102

\bibitem[{Ott {et~al.}(2004)Ott, Burrows, Livne, \& Walder}]{ott:2004:gwf}
Ott, C.~D., Burrows, A., Livne, E., \& Walder, R. 2004, ApJ, 600, 834

\bibitem[{{Ott} {et~al.}(2007){Ott}, {Dimmelmeier}, {Marek}, {Janka}, {Hawke},
  {Zink}, \& {Schnetter}}]{ott:2007:3co}
{Ott}, C.~D., {Dimmelmeier}, H., {Marek}, A., {Janka}, H.-T., {Hawke}, I.,
  {Zink}, B., \& {Schnetter}, E. 2007, Phys. Rev. Lett., 98, 261101

\bibitem[{Ott {et~al.}(2007)Ott, Dimmelmeier, Marek, Janka, Zink, Hawke, \&
  Schnetter}]{ott:2007:rco}
Ott, C.~D., Dimmelmeier, H., Marek, A., Janka, H.-T., Zink, B., Hawke, I., \&
  Schnetter, E. 2007, Class. Quantum Grav., 24, 139

\bibitem[{Pantin \& Starck(1996)}]{pantin:1996:doa}
Pantin, E., \& Starck, J.-L. 1996, Astron. Astrophys. Supp., 118, 575

\bibitem[{Ponsonby(1973)}]{ponsonby:1973:emf}
Ponsonby, J. E.~B. 1973, {MNRAS}, 163, 369

\bibitem[{Rakhmanov(2006)}]{rakhmanov:2006:rda}
Rakhmanov, M. 2006, Class. Quantum Grav., S673

\bibitem[{Rampp {et~al.}(1998)Rampp, Mueller, \& Ruffert}]{rampp:1998:son}
Rampp, M., Mueller, E., \& Ruffert, M. 1998, A\&A, 332, 969

\bibitem[{Robertson(2000)}]{robertson:2000:trl}
Robertson, N.~A. 2000, Class. Quantum Grav., 17, 19

\bibitem[{{R{\"u}diger}(2004)}]{rudiger:2004:l-l}
{R{\"u}diger}, A. 2004, in Observing, Thinking and Mining the Universe, ed.
  G.~{Miele} \& G.~{Longo} (Singapore: World Scientific Press), 329--+

\bibitem[{Shannon(1948{\natexlab{a}})}]{shannon:1948:mto}
Shannon, C.~E. 1948{\natexlab{a}}, The Bell System Technical Journal, 27, 327

\bibitem[{Shannon(1948{\natexlab{b}})}]{shannon:1948:mto:1}
---. 1948{\natexlab{b}}, The Bell System Technical Journal, 27, 623

\bibitem[{Shevgaonkar(1987)}]{shevgaonkar:1987:mem}
Shevgaonkar, R.~K. 1987, A\&A, 176, 159

\bibitem[{Shibata \& Sekiguchi(2005)}]{shibata:2005:tso}
Shibata, M., \& Sekiguchi, Y.-I. 2005, Phys. Rev. D, 71, 024014

\bibitem[{Skilling \& Bryan(1984)}]{skilling:1984:mei}
Skilling, J., \& Bryan, R.~K. 1984, {MNRAS}, 211, 111

\bibitem[{Starck \& Pantin(1996)}]{starck:1996:mme}
Starck, J., \& Pantin, E. 1996, Vistas in Astronomy, 40, 563

\bibitem[{Starck \& Pantin(1997)}]{starck:1997:air}
Starck, J. \& Pantin, E. 1997, in Statistical Challenges in Modern Astronomy
  II, ed. G.~J. {Babu} \& E.~D. {Feigelson} (Berlin: Springer-Verlag), 405

\bibitem[{Steenstrup(1985)}]{steenstrup:1985:dip}
Steenstrup, S. 1985, Australian Journal of Physics, 38, 319

\bibitem[{Takahashi \& {the TAMA Collaboration}(2004)}]{takahashi:2004:sot}
Takahashi, R., \& {the TAMA Collaboration}. 2004, Class. Quantum Grav., 21,
  S403

\bibitem[{Thorne(1987)}]{thorne:1987:gr}
Thorne, K.~S. 1987, in Three Hundred Years of Gravitation, ed. S.~Hawking \&
  W.~Israel (Cambridge, U.K.; New York, U.S.A.: Cambridge University Press),
  330--458

\bibitem[{Vielva {et~al.}(2001)Vielva, Barreiro, Hobson,
  {Mart{\'\i}nez-Gonz{\'a}lez}, Lasenby, Sanz, \& Toffolatti}]{vielva:2001:cma}
Vielva, P., Barreiro, R.~B., Hobson, M.~P., {Mart{\'\i}nez-Gonz{\'a}lez}, E.,
  Lasenby, A.~N., Sanz, J.~L., \& Toffolatti, L. 2001, {MNRAS}, 328, 1

\bibitem[{Waldman \& {The {LIGO} Scientific Collaboration
  }(2006)}]{waldman:2006:sol}
Waldman, S.~J., \& {The {LIGO} Scientific Collaboration }. 2006, Class. Quantum
  Grav., 23, S653

\bibitem[{Woosley \& Bloom(2006)}]{woosley:2006:sbc}
Woosley, S.~E., \& Bloom, J.~S. 2006, ARA\&A, 44, 508

\bibitem[{Woosley \& Weaver(1995)}]{woosley:1995:eae}
Woosley, S.~E., \& Weaver, T.~A. 1995, ApJS, 101

\end{thebibliography}

\clearpage

\begin{table}
\caption{Signal to noise $\rho^2$ and maximum cross-correlation $\max_j C(j)$ for a simulated gravitational wave signal incident on a two-detector LIGO-like network. The signal and detector network model are described in \S\ref{sec:Example} .}\label{tbl:example}
\begin{tabular}{ccc}
$h_0$&$\rho^2$&$\max_j C(j)$\\
\tableline\tableline
1&$0.29$&0.08\\
2&$1.5$&0.11\\
5&$1.5\times10^2$&0.54\\
10&$1.0\times10^{3}$&0.83\\
20&$5.4\times10^{3}$&0.94\\
\tableline
\end{tabular}
\end{table}

\clearpage

\end{document}